\documentclass[11pt]{amsart}
\usepackage{amsmath}
\usepackage{latexsym}
\usepackage{amsfonts}
\usepackage{amssymb}

\newcommand{\bm}[1]{\mb{\boldmath ${#1}$}}
\newcommand{\beqa}{\begin{eqnarray*}}
\newcommand{\eeqa}{\end{eqnarray*}}
\newcommand{\beqn}{\begin{eqnarray}}
\newcommand{\eeqn}{\end{eqnarray}}

\newcommand{\bs}{\boldsymbol}
\newcommand{\lt}{\left}
\newcommand{\rt}{\right}

\newcommand{\R}{\mathbb R}

\newcommand{\mcL}{\mathcal L}

\newcommand{\mb}{\makebox}

\newcommand{\f}{\frac}
\newcommand{\tf}{\tfrac}

\newcommand{\al}{\alpha}
\newcommand{\be}{\beta}

\newcommand{\e}{\varepsilon}

\newcommand{\de}{\delta}
\newcommand{\De}{\Delta}
\newcommand{\la}{\lambda}

\newcommand{\s}{\sigma}

\newcounter{cnt1}
\newcounter{cnt2}
\newcounter{cnt3}
\newcommand{\blr}{\begin{list}{$($\roman{cnt1}$)$}
 {\usecounter{cnt1} \setlength{\topsep}{0pt}
 \setlength{\itemsep}{0pt}}}
\newcommand{\bla}{\begin{list}{$($\alph{cnt2}$)$}
 {\usecounter{cnt2} \setlength{\topsep}{0pt}
 \setlength{\itemsep}{0pt}}}
\newcommand{\bln}{\begin{list}{$($\arabic{cnt3}$)$}
 {\usecounter{cnt3} \setlength{\topsep}{0pt}
 \setlength{\itemsep}{0pt}}}
\newcommand{\el}{\end{list}}
\newtheorem{thm}{Theorem}[section]

\newtheorem{Def}[thm]{Definition}

\newtheorem{rem}[thm]{Remark}
\newcommand{\Rem}{\begin{rem} \rm}
\newcommand{\bdfn}{\begin{Def} \rm}
\newcommand{\edfn}{\end{Def}}

\newcommand{\ba}{\begin{array}}
\newcommand{\ea}{\end{array}}

\numberwithin{equation}{section}
\date{}
\begin{document}
\title{\bf Foundations for Proper-time Relativistic Quantum Theory}
\author[Gill]{T. L. Gill}
\address[Tepper L. Gill]{ Department of Mathematics, Physics and E\&CE, 
Howard University\\ Washington DC 20059 \\ USA, {\it E-mail~:} {\tt tgill@howard.edu}}
\author[Morris]{T.  Morris}
\address[Trey  Morris]{ Department of E\&CE, Howard University \\
Washington DC 20059 \\ USA, {\it E-mail~:} {\tt Morris.Trey.J@gmail.com}}
\author[Kurtz]{S. K. Kurtz}
\address[Stewart K. Kurtz]{ Department of Electrical Engineering, Penn. State University \\
University Park PA 16802-2703 \\ USA, {\it E-mail~:} {\tt  skk1@psu.edu}}
\date{}
\keywords{ Electrodynamics · Proper-time · Dirac Equation · QED }
\maketitle
\begin{abstract}  This paper is  a progress report on the foundations for the canonical proper-time  approach to relativistic quantum theory.  We first review the the standard square-root equation of relativistic quantum theory, followed by a review of the Dirac equation, providing  new insights into the physical properties of both.  We then introduce the canonical proper-time theory.  For completeness, we give a brief outline of the canonical proper-time approach to electrodynamics and mechanics, and then introduce the canonical proper-time approach to relativistic quantum theory.  This theory leads to three new relativistic wave equations.  In each case, the canonical generator of proper-time translations is strictly positive definite, so that each represents a true particle equation.  We show that the canonical proper-time version of the  Dirac equation for Hydrogen gives results that are consistently closer to the experimental data, when compared to the Dirac equation. However, these results are not sufficient to account for either the Lamb shift or the anomalous magnetic moment. 
\end{abstract}
\footnote{This paper is an extended version of the talk given at the The 9th Biennial Conference on Classical and Quantum Relativistic Dynamics of Particles and Fields held at the University of Connecticut in June 2014.}
\tableofcontents
\section*{Introduction}
Following Dirac's quantization of the electromagnetic field in 1927, and his relativistic electron theory in 1928,  the equations for quantum electrodynamics QED were developed by Heisenberg and Pauli in the years 1929-30. From the beginning, when researchers attempted to use the straightforward and physically intuitive time-dependent perturbation expansion to compute physical observables, a number of divergent expressions appeared. Although it was known that the same problems also existed in classical electrodynamics, Dirac had shown that, in this case, one could account for the problem of radiation reaction without directly dealing with the self-energy divergence by using both advanced and retarded fields and a particular limiting procedure.  Early attempts to develop subtraction procedures for the divergent expressions were very discouraging because they depended on both the gauge and the Lorentz frame, making them appear ambiguous. These problems were solved via the fundamental work of Feynman, Schwinger, and Tomonaga. In recent times, it is generally agreed that quantum electrodynamics (QED) is an almost perfect theory, which is in excellent agreement with experiment. 

The fact that QED is very successful is without doubt. However, there are still some foundational and technical issues, which require clarification and which leaves the thoughtful student with a sense of unease in taking this as the final answer.  In light of the tremendous historical success of eigenvalue analysis in physics and engineering, it is not inappropriate  to reinvestigate the foundations with an eye towards clearly identifying the physical and mathematical limitations to our understanding of the hydrogen spectrum as an eigenvalue problem.

In the first section of this review we take a new look at the square-root operator and show that it has an analytic  representation as a nonlocal composite of three singularities. The particle component has two negative parts and one (hard core) positive part, while the antiparticle component has two positive parts and one (hard core) negative part. This effect is confined within a Compton wavelength such that, at the point of singularity, they cancel each other providing a finite result.  Furthermore, the operator looks like the identity outside a  Compton wavelength.

In the second section, we provide an analytic diagonalization of the Dirac operator.  Our approach leads to a complete split of the particle and antiparticle parts into two non-hemitian components, which are mapped into each other by the charge conjugation transformation. Thus, the full matrix-valued operator is Hermitian and shows (as is explained in the text) that the spinor representation in the Dirac equation hides its time nonlocal property.  We conclude that the Dirac and square-root operator do not represent the same physics, despite that fact that they are related by a unitary transformation. 

In the third section, we introduce the canonical proper-time approach to electrodynamics and mechanics.  By convention, this approach fixes the proper-time of the observed system as the clock of choice for all observers and explicitly shows that the question of simultaneity is actually a question of clock conventions.  The change in convention produces a new symmetry group which is distinct from, but closely related to the Lorentz group, but with a Euclidean representation space. Thus, the new convention also replaces the standard form of Lorentz covariance by a new one.  The advantage is that, this allows us to construct a parallel image of the conventional Maxwell theory for a charged particle, which is mathematically, but not physically, equivalent to the conventional form.  The new wave equation contains a gauge independent term, which appears instantaneously along the direction of motion, but opposing  any applied force and is zero otherwise.  This is the near field (i.e., the field at the site of the charged particle). This shows that the origin of radiation reaction is not the action of a charge on itself but arises from inertial resistance to changes in motion. We show that the dissipative term is equivalent to an effective mass so that classical radiation has both a massless and a massive part. We also discuss solutions to a number of other problems that are solved with our new convention, which are either impossible or problematic within the standard framework.

In the forth section, we describe the canonical quantized proper-time theory.  We obtain three possible relativistic wave equations,  because of new possibilities, for the manner in which the potential energy may be introduced into the theory.  Each new equation is generated by a strictly positive definite canonical Hamiltonian, so that it represents a consistent particle. We focus on the proper-time extension of the Dirac equation.  A basic test of our proper-time theory is the extent that it compares to the Dirac theory  in accounting for the hydrogen spectra.  We show that our canonical proper-time version of the  Dirac equation gives results which are consistently closer to the experimental data, when compared to the Dirac equation.   The  present theory has not yet accounted for the  Lamb shift or the anomalous magnetic moment.  However, the analysis in sections one, two and three support our contention that the electron is not a point particle.  This non-point nature is only expected to be important in s-states, where there is a finite probability of the electron being at the center of the proton.  This aspect of our research is still in progress and will be reported on at a later time.
\section{The square-root equation}
In the transition to  relativistic mechanics, the equation $E^2 =c^2{\bf{p}}^2 + {m^2 c^4}$ leads to the quantum Hamiltonian 
\[
H = \sqrt {c^2 {\mathbf{p}}^2 + m^2 c^4}. 
\]
Thus, it is quite natural to expect that the first choice for a relativistic wave equation would be:
\[
i\hbar \frac{{\partial \psi }}{{\partial t}} = \left[ {\sqrt {{c^2}{{\left[ {{\mathbf{p}} - {{(e} \mathord{\left/
 {\vphantom {{(e} c}} \right. \kern-\nulldelimiterspace} c}){\mathbf{A}}} \right]}^2} + {m^2}{c^4}}  + V} \right]\psi,
 \]
where ${\mathbf{p}} =  - i\hbar \nabla $.  However, no one knew how to directly relate this equation to physically important problems.  Furthermore, this equation is nonlocal, meaning, in the terminology of the times (1920-30), that it is represented by a power series in the momentum operator.  One was led in this way to the Gordon-Klein and Dirac equations.
\subsection{Background} Since the early work, many investigators have studied the square-root equation.  It is not our intention to provide a detailed history or to identify the many important contributors to the study of this problem.  In recent times, the works of Silenko (see \cite{1}, \cite{2}) are well worth reading.  They also provide a very good list of the important historical studies.  In addition, he has made a number of interesting investigations into the transformational relationship between the square-root and Dirac equation (\cite{3} is a good starting point).   The recent paper by Simulik and Krivsky \cite{4} offers another interesting approach to the square-root equation and its relationship to that of Dirac.   Closer to our investigation of the square-root equation is the study by Kowalski and Rembieli\'{n}ski \cite{5} (also known as the Salpeter equation). They have used it as an alternative of the Klein-Gordon equation.   

In this section, we take a new look at the square-root equation.  First, we investigate the extent that the non-commutativity of ${\mathbf{p}}$  and ${\mathbf{A}}$  affect our ability to give an unambiguous meaning to the square-root operator.  We show that a unique analytic representation is well defined for suitable time-independent ${\mathbf{A}}$ provided we can solve a corresponding equation of the Schr\"{o}dinger type.  We then investigate a few simple cases of solvable models in order to get a feeling for the physical interpretation of this operator.    

To begin, we start with the equation: 
\beqn
S[\psi ] = {{{H}}_s}\psi  = \left\{ {\beta \sqrt {{c^2}{{\left( {{\mathbf{p}} - \tfrac{e}{c}{\mathbf{A}}} \right)}^2} - e\hbar c\Sigma  \cdot {\mathbf{B}} + {m^2}{c^4}} } \right\}\psi. 
\eeqn
Where $\beta$ and $\Sigma $ are the Dirac matrices
\[
\beta  = \left[ {\begin{array}{*{20}{c}}
  {\mathbf{I}}& \bm{0} \\ 
  \bm{0}&{ - {\mathbf{I}}} 
\end{array}} \right],\;{\text{  }}\Sigma  = \left[ {\begin{array}{*{20}{c}}
  \bm{\sigma} &\bm{0} \\ 
  \bm{0}&\bm{\sigma}  
\end{array}} \right];
\]
$\bf{I}$ and $\bm{\s}$ are the identity and Pauli matrices respectively.  Under physically reasonable mathematical conditions, the following operator is a well defined self-adjoint generator of a strongly continuous unitary group: 
\[
H_s^2 ={{c^2}{{\left[ {{\mathbf{p}} - {{(e} \mathord{\left/
 {\vphantom {{(e} c}} \right. \kern-\nulldelimiterspace} c}){\mathbf{A}}} \right]}^2} - e\hbar c\Sigma  \cdot {\mathbf{B}}+ {m^2}{c^4}}.
\]
From the basic theory of fractional powers of closed linear operators, it can be shown that
\beqn
\sqrt {{ H}_s^2}  = {\left( {\sqrt {{ H}_s^2} } \right)^{ - 1}}{ H}_s^2 = { H}_s^2{\left( {\sqrt {{ H}_s^2} } \right)^{ - 1}}.
\eeqn
In order to construct an analytic representation for equation (1.1), we assume that $\bf{B}$ is constant.  The general case can be found in \cite{6}. Let ${\mathbf{G}} =  - c^2{\left( {{\mathbf{p}} - \tfrac{e}{c}{\mathbf{A}}} \right)^2}$ and ${\omega ^2} = {m^2}{c^4} - {{e\hbar }}{c}\Sigma  \cdot {\mathbf{B}}$, so that $\omega$ is also constant.  Using  this notation, we can write (1.1) as
\[
S[\psi ] = \left\{ {\beta \sqrt { - {\mathbf{G}} + {\omega ^2}} } \right\}\psi. 
\]
Using the analytic theory of fractional powers of closed linear operators and equation (1.2), we can represent $S[\psi]$ as 
\beqn
S[\psi ] = \frac{{\beta }}{\pi }\int_0^\infty  {{{\left[ {(\lambda  + {\omega ^2}) - {\mathbf{G}}} \right]}^{ - 1}}( - {\mathbf{G}} + {\omega ^2})\frac{{d\lambda }}{{\sqrt \lambda  }}\left[ \psi  \right]} ,
\eeqn
where $\left[ {(\lambda  + {\omega ^2}) - {\mathbf{G}}} \right]^{ - 1}$ is the resolvent associated with the operator ${ - {\mathbf{G}} + {\omega ^2}}$. The resolvent can be computed directly if we can find the fundamental solution to the equation:
\[
{{\partial Q({\mathbf{x}},{\mathbf{y}};t)} \mathord{\left/
 {\vphantom {{\partial Q({\mathbf{x}},{\mathbf{y}};t)} {\partial t}}} \right.
 \kern-\nulldelimiterspace} {\partial t}} + ({\mathbf{G}} - {\omega ^2})Q({\mathbf{x}},{\mathbf{y}};t) = \delta ({\mathbf{x}} - {\mathbf{y}}).
\]
Schulman \cite{7} has shown that the solution to the above equation is
\[
Q = \int_{{\mathbf{x}}(0)}^{{\mathbf{x}}(t)} {{\mathcal{D}}[{\mathbf{x}}(s)]} \exp \left\{ {\int_0^t {V[{\mathbf{x}}(s)]ds}  + \tfrac{{ie}}{{\hbar c}}\int_{\mathbf{y}}^{\mathbf{x}} {{\mathbf{A}}[{\mathbf{x}}(s)]}  \cdot d{\mathbf{x}}(s)} \right\}
\]
where $V = {c^2\omega ^2}/i\hbar$ and
\[
\begin{gathered}
  \int_{{\mathbf{x}}(0) = {\mathbf{y}}}^{{\mathbf{x}}(t) = {\mathbf{x}}} {{\mathcal{D}}{{}[{\mathbf{x}}(s)]}  = \int_{{\mathbf{x}}(0) = {\mathbf{y}}}^{{\mathbf{x}}(t) = {\mathbf{x}}} {{\mathcal{D}}}[{\mathbf{x}}(s)]} \exp \left\{ { - \tfrac{1}{4}\int_0^t {{{\left| {\frac{{d{\mathbf{x}}(s)}}{{ds}}} \right|}^2}} ds} \right\} \hfill \\
  \quad  = \mathop {\lim }\limits_{N \to \infty } {\left[ {\tfrac{1}{{4\pi {\varepsilon _N}}}} \right]^{nN/2}}\mathop \smallint \nolimits_{{{\mathbb{R}}^n}} \mathop \prod \limits_{k = 1}^N \mathop {dx}\nolimits_j \exp \left\{ { - \mathop \sum \limits_{j = 1}^N \left[ {\tfrac{1}{{4{\varepsilon _N}}}\mathop {\left( {\mathop x\nolimits_j  - \mathop x\nolimits_{j - 1} } \right)}\nolimits^2 } \right]} \right\}, \hfill \\ 
\end{gathered} 
\]
and $\e_N=t/N$.  A rigorous justification for the path integral can be found in Gill and Zachary \cite{8}.  We assume that $\int_{\mathbf{y}}^{\mathbf{x}} {{\mathbf{A}}[{\mathbf{x}}(s)]}  \cdot d{\mathbf{x}}(s) = {\mathbf{\bar A}} \cdot ({\mathbf{x}} - {\mathbf{y}})$, where ${\mathbf{\bar A}}$ is the mean value of ${\mathbf{ A}}$.  Using this, we have:
\[
{\left[ {(\lambda  + {\omega ^2}) - {\mathbf{G}}} \right]^{ - 1}}f({\mathbf{x}}) = \int_0^\infty  {{e^{ - \lambda t}}} \left[ {\int_{{\mathbb{R}^3}} {Q({\mathbf{x}},t;{\mathbf{y}},0)f({\mathbf{y}})d{\mathbf{y}}} } \right]dt
\]
and
\[
\begin{gathered}
  {\left[ {(\lambda  + {\omega ^2}) - {\mathbf{G}}} \right]^{ - 1}}f({\mathbf{x}}) \hfill \\
  {\text{  }} = \int_{{\mathbb{R}^3}} {e^{\left\{ {\tfrac{{ie}}{{\hbar c}}{\mathbf{\bar A}} \cdot \left( {{\mathbf{x}} - {\mathbf{y}}} \right)} \right\}}} \left\{ {\int_0^\infty  {\exp \left[ { - \tfrac{{{{\left( {{\mathbf{x}} - {\mathbf{y}}} \right)}^2}}}{{4t}} - \tfrac{{{\omega ^2}t}}{{{\hbar ^2}}} - \lambda t} \right]} \tfrac{{dt}}{{{{\left( {4\pi t} \right)}^{3/2}}}}} \right\}f({\mathbf{y}})d{\mathbf{y}} \hfill \\ 
\end{gathered} 
\]
Using a table of Laplace transforms, the inner integral can be computed to get
\[
\begin{gathered}
  \int_0^\infty  {\exp \left[ { - \frac{{{{\left( {{\mathbf{x}} - {\mathbf{y}}} \right)}^2}}}{{4t}} - \frac{{{\omega ^2}t}}{{{\hbar ^2}}} - \lambda t} \right]} \frac{{dt}}{{{{\left( {4\pi t} \right)}^{3/2}}}} \hfill \\
   = \frac{1}{{4\pi }}\frac{{\exp \left[ { - \sqrt {\left( {\lambda  + {\mu ^2}} \right)} \left\| {{\mathbf{x}} - {\mathbf{y}}} \right\|} \right]}}{{\left\| {{\mathbf{x}} - {\mathbf{y}}} \right\|}} \hfill \\ 
\end{gathered} 
\]
where ${\mu ^2} = {\omega ^2}/{\hbar ^2}$.  Equation (1.3) now becomes
\[
\begin{gathered}
  S[\psi ]({\mathbf{x}}) \hfill \\
   = \tfrac{{c\beta }}{{4{\pi ^2}}}\int_0^\infty  {\left\{ {\int_{{\mathbb{R}^3}} {{\operatorname{e} ^{\left\{ {\tfrac{{ie}}{{\hbar c}}{\mathbf{\bar A}} \cdot \left( {{\mathbf{x}} - {\mathbf{y}}} \right)} \right\}}}} {\operatorname{e} ^{ - \left[ {\sqrt {\left( {\lambda  + {\mu ^2}} \right)} \left\| {{\mathbf{x}} - {\mathbf{y}}} \right\|} \right]}}\tfrac{{( - {\mathbf{G}} + {\omega ^2})}}{{\left\| {{\mathbf{x}} - {\mathbf{y}}} \right\|}}\psi ({\mathbf{y}})d{\mathbf{y}}} \right\}\frac{{d\lambda }}{{\sqrt \lambda  }}}.  \hfill \\ 
\end{gathered} 
\]
Once again, we interchange the order of integration and perform the computations to get 
\[
\int_0^\infty  {\left\{ {\frac{{\exp \left[ { - \sqrt {\left( {\lambda  + {\mu ^2}} \right)} \left\| {{\mathbf{x}} - {\mathbf{y}}} \right\|} \right]}}{{\left\| {{\mathbf{x}} - {\mathbf{y}}} \right\|}}} \right\}\frac{{d\lambda }}{{\sqrt \lambda  }} = \frac{{4\mu \Gamma (\tfrac{3}{2})}}{{{\pi ^{1/2}}}}\frac{{{{\mathbf{K}}_1}\left[ {\mu \left\| {{\mathbf{x}} - {\mathbf{y}}} \right\|} \right]}}{{\left\| {{\mathbf{x}} - {\mathbf{y}}} \right\|}}} .
\]
where ${\mathbf{K}}_1[\bf{z}]$ is the modified Bessel function of the third kind and first order.
Thus, if we set, ${\mathbf{a}} = \tfrac{e}{{\hbar c}}{\mathbf{A}}$ and ${\mathbf{\bar a}} = \tfrac{e}{{\hbar c}}{\mathbf{\bar A}}$ we get
\beqn
\begin{gathered}
  S[\psi ]({\mathbf{x}}) = \tfrac{{c\beta }}{{2{\pi ^2}}}\int_{{{\mathbf{R}}^3}} {{\operatorname{e} ^{\left[ {i{\mathbf{\bar a}} \cdot \left( {{\mathbf{x}} - {\mathbf{y}}} \right)} \right]}}\frac{{\mu {{\mathbf{K}}_1}\left[ {\mu \left\| {{\mathbf{x}} - {\mathbf{y}}} \right\|} \right]}}{{\left\| {{\mathbf{x}} - {\mathbf{y}}} \right\|}}( - {\mathbf{G}} + {\omega ^2})\psi ({\mathbf{y}})} d{\mathbf{y}} \hfill \\
   = \tfrac{{c\beta }}{{2{\pi ^2}}}( - {\mathbf{G}} + {\omega ^2})\int_{{{\mathbf{R}}^3}} {{\operatorname{e} ^{\left[ {i{\mathbf{\bar a}} \cdot \left( {{\mathbf{x}} - {\mathbf{y}}} \right)} \right]}}\frac{{\mu {{\mathbf{K}}_1}\left[ {\mu \left\| {{\mathbf{x}} - {\mathbf{y}}} \right\|} \right]}}{{\left\| {{\mathbf{x}} - {\mathbf{y}}} \right\|}}\psi ({\mathbf{y}})} d{\mathbf{y}}. \hfill \\ 
\end{gathered} 
\eeqn
Since $\nabla  \cdot {\mathbf{a}}=0$, we have
\[ 
- {\mathbf{G}} + {\omega ^2} = {\hbar ^2}\left( { - \Delta  + 2i{\mathbf{a}} \cdot \nabla  + {{\mathbf{a}}^2} + {\mu ^2}} \right),
\]
so that so that (1.4) becomes
\[
S[\psi ]({\mathbf{x}}) = \tfrac{{{\hbar ^2}c\beta }}{{2{\pi ^2}}} \left( { - \Delta  + 2i{\mathbf{a}} \cdot \nabla  + {{\mathbf{a}}^2} + {\mu ^2}} \right) \int_{{{\mathbf{R}}^3}} {{e^{i{\mathbf{\bar a}} \cdot ({\mathbf{x}} - {\mathbf{y}})}}\tfrac{{\mu {{\mathbf{K}}_1}\left[ {\mu \left\| {{\mathbf{x}} - {\mathbf{y}}} \right\|} \right]}}{{\left\| {{\mathbf{x}} - {\mathbf{y}}} \right\|}}\psi ({\mathbf{y}})} d{\mathbf{y}}.
\]
The operator $\left( { - \Delta  + 2i{\mathbf{a}} \cdot \nabla  + {{\mathbf{a}}^2} + {\mu ^2}} \right)$ acts on $\bf{x}$, making the integral singular.  However, this singular representation constructed below, equation (1.5) has many of the properties observed in experiments.  As will be seen, it represents the confinement of three singularities within a Compton wavelength.  (A full discussion is delayed to the end of this section.)

We omit many of the computational details, which can be found in \cite{9}, but the idea is to consider a ball ${{\mathbf{B}}_\rho }({\mathbf{x}})$ of radius $\rho$ about $\bf{x}$, so that ${\mathbb{R}^3} = \mathbb{R}_\rho ^3 \cup {{\mathbf{B}}_\rho }({\mathbf{x}})$, where $\mathbb{R}_\rho ^3 = \left( {{\mathbb{R}^3}\backslash {{\mathbf{B}}_\rho }({\mathbf{x}})} \right)$ and $\partial \mathbb{R}_\rho ^3 = \left( {\partial {\mathbb{R}^3}\backslash \partial {{\mathbf{B}}_\rho }({\mathbf{x}})} \right)$.  We then restrict all operations to $\R_\rho^3$ and only let $\rho \to 0$ at the end.
\subsection{Free Case}
The free particle case is the simplest (but still interesting), with $\bf{A}=0$, so that
\beqn
\begin{gathered}
  S[\psi ]({\mathbf{x}}) \hfill \\
   =  - \tfrac{{{\mu ^2}{\hbar ^2}c\beta }}{{{\pi ^2}}}\int\limits_{{{\mathbf{R}}^3}} {\left[ {\tfrac{1}{{\left\| {{\mathbf{x}} - {\mathbf{y}}} \right\|}} - 4\pi \delta \left( {{\mathbf{x}} - {\mathbf{y}}} \right)} \right]\left\{ {\tfrac{{{{\mathbf{K}}_0}\left[ {\mu \left\| {{\mathbf{x}} - {\mathbf{y}}} \right\|\,} \right]}}{{\,\left\| {{\mathbf{x}} - {\mathbf{y}}} \right\|}} + \tfrac{{2{{\mathbf{K}}_1}\left[ {\mu \left\| {{\mathbf{x}} - {\mathbf{y}}} \right\|\,} \right]}}{{\mu \,{{\left\| {{\mathbf{x}} - {\mathbf{y}}} \right\|}^2}}}} \right\}\psi ({\mathbf{y}})d{\mathbf{y}}} . \hfill \\ 
\end{gathered} 
\eeqn
If $\bf{x} \ne \bf{y}$, the effective kernel of equation (1.5) is 
\[
\frac{{{K_0}\left[ {\mu \left\| {{\mathbf{x}} - {\mathbf{y}}} \right\|} \right]}}{{{{\left\| {{\mathbf{x}} - {\mathbf{y}}} \right\|}^2}}} + \frac{{2{K_1}\left[ {\mu \left\| {{\mathbf{x}} - {\mathbf{y}}} \right\|} \right]}}{{\mu {{\left\| {{\mathbf{x}} - {\mathbf{y}}} \right\|}^3}}}.
\]
Recall that the integral of ${\left\| {{\mathbf{x}} - {\mathbf{y}}} \right\|^{ - 2}}$ over $\R^3$ is finite.
In order to understand the physical interpretation of equation (1.5), it will be helpful to review some properties of the modified Bessel functions $K_0[u]$, \, $u^{-\tf{1}{2}}K_{{1}/{2}}[u]$ and $u^{-1}K_1[u]$.  We follow Gradshteyn and Ryzhik \cite{10}, for $0 < u \ll 1$, we have that:
\[
\begin{gathered}
  \frac{{{K_1}\left[ u \right]}}{u} = \left[ {1 + {\theta _1}(u)} \right]{u^{ - 2}} \hfill \\
  \frac{{{K_{1/2}}\left[ u \right]}}{{{u^{1/2}}}} = \sqrt {\tfrac{\pi }{2}} {u^{ - 1}} \hfill \\
  {K_0}\left[ u \right] = \left[ {1 + {\theta _0}(u)} \right]\ln {u^{ - 1}}, \hfill \\ 
\end{gathered} 
\]
where $\theta _0, \, \theta _1 \to 0$ as $u \to 0$.  We note that, up to a multiplicative constant, $u^{-\tf{1}{2}}K_{{1}/{2}}[u]$ is the well-known Yukawa potential \cite{11}, conjectured in 1935 to account for the short range of the nuclear interaction.  From here, we see that, near $u=0$, the singular term $u^{-1}K_1[u]$ is twice as strong as the Yukawa potential.  The singular term $K_0[u]$ is actually integrable and so does not contribute at $u=0$.  Looking at equation (1.5), we see that the singular term $-8\pi K_1(u) \de(u)$ acts to cancel the the singular term $u^{-1}K_1[u]$ at $u=0$, so that the total integral is well defined.

The behavior of these functions is quite different, when  $u \gg 1$.  In this case, we have:
\[
\begin{gathered}
  \frac{{{K_1}\left[ u \right]}}{u} = \left[ {1 + \theta {'_1}(u)} \right]\frac{{\exp \left\{ { - u} \right\}}}{{{u^{3/2}}}} \hfill \\
  \frac{{{K_{1/2}}\left[ u \right]}}{{{u^{1/2}}}} = \sqrt {\tfrac{\pi }{2}} \frac{{\exp \left\{ { - u} \right\}}}{u} \hfill \\
  {K_0}\left[ u \right] = \left[ {1 + \theta {'_0}(u)} \right]\frac{{\exp \left\{ { - u} \right\}}}{{{u^{1/2}}}}. \hfill \\ 
\end{gathered} 
\] 
We see that each term has a exponential cutoff.  However, now  $K_0[u]$ has the longest range, while $u^{-1}K_1[u]$ has the shortest range.  Furthermore, inspection shows that $u^{-1}K_1[u]$ is multiplied by the reduced Compton wavelength, which further shortens its  range.

It is clear that our square-root operator represents an extended object with an effective extent of about a Compton wavelength.
\subsection{Constant $\bf{A}$ Case}
When $\bf{A} \ne 0$ is constant, $\nabla  \cdot {\mathbf{A}} = 0$ and ${\mathbf{B}} = \nabla  \times {\mathbf{A}} = 0$, so that we get (${\mathbf{\bar a}} = {\mathbf{a}}$):
\[
\begin{gathered}
  S[\psi ]({\mathbf{x}})    =  \hfill \\
\tfrac{{{- \mu ^2}{\hbar ^2}c\beta }}{{{\pi ^2}}}\int\limits_{{{\mathbf{R}}^3}} {{e^{i{\mathbf{a}} \cdot ({\mathbf{x}} - {\mathbf{y}})}}\left[ {\tfrac{1}{{\left\| {{\mathbf{x}} - {\mathbf{y}}} \right\|}} - 4\pi \delta \left( {{\mathbf{x}} - {\mathbf{y}}} \right)} \right]\left\{ {\tfrac{{{{\mathbf{K}}_0}\left[ {\mu \left\| {{\mathbf{x}} - {\mathbf{y}}} \right\|\,} \right]}}{{\,\left\| {{\mathbf{x}} - {\mathbf{y}}} \right\|}} + \tfrac{{2{{\mathbf{K}}_1}\left[ {\mu \left\| {{\mathbf{x}} - {\mathbf{y}}} \right\|\,} \right]}}{{\mu \,{{\left\| {{\mathbf{x}} - {\mathbf{y}}} \right\|}^2}}}} \right\}\psi ({\mathbf{y}})d{\mathbf{y}}} . \hfill \\ 
\end{gathered} 
\]
\subsection{The Constant Field Case}
If $\bf{B} \ne 0$ is constant, then ${\mathbf{A}}({\mathbf{z}}) = \tfrac{1}{2}{\mathbf{z}} \times {\mathbf{B}}$.  Let ${\mathbf{a}}({\mathbf{z}}) = \tfrac{e}{2\hbar c}{\mathbf{z}} \times {\mathbf{B}}$ and   $F = - {\mathbf{a}} \cdot ({\mathbf{x}} - {\mathbf{y}})$.   In this case $\bf{A}({\bf{z}}) \cdot d{\bf{z}}=0$, so we can write the final result as:
\beqn
\begin{gathered}
  {\mathbf{S}}[\psi ] =  - \tfrac{{{\hbar ^2}{\mu ^2}c\beta }}{{{\pi ^2}}}\left\{ {\int_{{{\mathbf{R}}^3}}  \left[ {\tfrac{1}{{\left\| {{\mathbf{x}} - {\mathbf{y}}} \right\|}} - \tfrac{{4\pi \delta \left( {{\mathbf{x}} - {\mathbf{y}}} \right)}}{{1 + iF}}} \right]\left[ {1 + iF} \right]} \right.\left. {\tfrac{{{{\mathbf{K}}_2}\left[ {\mu \left\| {{\mathbf{x}} - {\mathbf{y}}} \right\|} \right]}}{{\left\| {{\mathbf{x}} - {\mathbf{y}}} \right\|}}\psi ({\mathbf{y}})d{\mathbf{y}}} \right\} \hfill \\
  {\text{         }} + \tfrac{{{\hbar ^2}{\mu ^2}c\beta }}{{{\pi ^2}}}\int_{{{\mathbf{R}}^3}}  {{{\mathbf{a}}^2}  \tfrac{{{{\mathbf{K}}_1}\left[ {\mu \left\| {{\mathbf{x}} - {\mathbf{y}}} \right\|} \right]}}{{\left\| {{\mathbf{x}} - {\mathbf{y}}} \right\|}}\psi ({\mathbf{y}}) d{\mathbf{y}}}, \hfill \\ 
\end{gathered} 
\eeqn
where
\beqn
\frac{{{{\mathbf{K}}_2}\left[ {\mu \left\| {{\mathbf{x}} - {\mathbf{y}}} \right\|} \right]}}{{\left\| {{\mathbf{x}} - {\mathbf{y}}} \right\|}} = \frac{{{{\mathbf{K}}_0}\left[ {\mu \left\| {{\mathbf{x}} - {\mathbf{y}}} \right\|} \right]}}{{\left\| {{\mathbf{x}} - {\mathbf{y}}} \right\|}} + \frac{{2{{\mathbf{K}}_1}\left[ {\mu \left\| {{\mathbf{x}} - {\mathbf{y}}} \right\|} \right]}}{{\mu {{\left\| {{\mathbf{x}} - {\mathbf{y}}} \right\|}^2}}}.
\eeqn
From equations (1.6) and (1.7), we see that a constant magnetic field makes a real difference compared to either the $\bf{A} = 0$ or $\bf{A} \ne 0$ cases, producing two extra terms, in addition to the free particle term.  The first new term is purely imaginary and singular at  $\bf{x}  =\bf{y}$ (like the Yukawa term).  Physically, we interpret this term as representing particle absorption and emission (see Mott  and Massey \cite{12}) .   The second term is real,  repulsive and nonsingular.   In addition, the effective mass  $\mu$ is constant but matrix-valued with complex components, 
${\mu ^2} = {{{m^2}{c^2}} \mathord{\left/
 {\vphantom {{{m^2}{c^2}} {{\hbar ^2}}}} \right.
 \kern-\nulldelimiterspace} {{\hbar ^2}}} - \tfrac{e}{{\hbar c}}\Sigma  \cdot {\mathbf{B}}$. Since
\[
\begin{gathered}
  \Sigma  = \left( {\begin{array}{*{20}{c}}
  \sigma &{\mathbf{0}} \\ 
  {\mathbf{0}}&\sigma  
\end{array}} \right);{\text{   }}{\sigma _1} = \left( {\begin{array}{*{20}{c}}
  {\mathbf{0}}&1 \\ 
  1&{\mathbf{0}} 
\end{array}} \right),{\text{ }}{\sigma _2} = \left( {\begin{array}{*{20}{c}}
  {\mathbf{0}}&{ - i} \\ 
  i&{\mathbf{0}} 
\end{array}} \right),{\text{ }}{\sigma _3} = \left( {\begin{array}{*{20}{c}}
  {\mathbf{1}}&{\mathbf{0}} \\ 
  {\mathbf{0}}&{ - {\mathbf{1}}} 
\end{array}} \right), \hfill \\
  {\mu ^2} = \left[ {\begin{array}{*{20}{c}}
  {(\tfrac{{{m^2}{c^2}}}{{{\hbar ^2}}} - \tfrac{e}{{\hbar c}}{B_3}){{\mathbf{I}}_2}}&{\tfrac{{ie}}{{\hbar c}}({B_2} - i{B_1}){{\mathbf{I}}_2}} \\ 
  {\tfrac{{ - ie}}{{\hbar c}}({B_2} - i{B_1}){{\mathbf{I}}_2}}&{(\tfrac{{{m^2}{c^2}}}{{{\hbar ^2}}} + \tfrac{e}{{\hbar c}}{B_3}){{\mathbf{I}}_2}} 
\end{array}} \right]. \hfill \\ 
\end{gathered} 
\]
From known properties of Bessel functions for nonintegral  $\nu$, we can represent  ${{\mathbf{K}}_\nu }[u]$ as
\[
({2 \mathord{\left/
 {\vphantom {2 \pi }} \right.
 \kern-\nulldelimiterspace} \pi }){{\mathbf{K}}_\nu }[u] = \frac{{{{\mathbf{I}}_{ - \nu }}(u) - {{\mathbf{I}}_\nu }(u)}}{{\sin \pi \nu }} = \frac{{{e^{i/2(\pi \nu )}}{{\mathbf{J}}_{ - \nu }}(iu) - {e^{ - i/2(\pi \nu )}}{{\mathbf{J}}_\nu }(iu)}}{{\sin \pi \nu }}.
\]
In the limit as  $\nu$ approaches an integer, the above takes the indeterminate form $0/0$ and is defined via L'H\^{o}pital's rule.  However, for our purposes, we assume that  $\nu$ is close to an integer 
and $u=u_1+iu_2,  \; u_2 \ne 0$.  In this case, ${{\mathbf{K}}_\nu }[u]$ acquires some of the oscillatory behavior of  ${{\mathbf{J}}_\nu }[u]$.    Thus, we can interpret equation (1.6) as representing a  pulsating mass (extended object of variable mass) with mean value $\hbar/c  \lt\| \mu \rt\|$, where  $\left\| \mu  \right\| = {\left[ {{\mu ^ * }\mu } \right]^{1/2}}$ and $\mu^*$  is the Hermitian conjugate of $\mu$,  with the square root being computed using elementary spectral theory.  If  $\bf{B}$ is very large, we see that the effective mass can also be large.  However, the operator still looks (almost) like the identity outside a Compton wavelength.  

In closing, we should say a few additional words about the interesting work of  Kowalski and Rembieli\'{n}ski \cite{5}.  They solve the  Salpeter equation ($\be =I$) and construct a number of examples.  This work represents an original contribution to our understanding of the square-root equation.  They approach the problem using the method of Fourier transforms and get the correct solution for $\bf{x} \ne \bf{y}$.  However, this approach misses the the $\bf{x} = \bf{y}$ term, giving the impression that their equation is not defined at that point.  This minor defect  can be easily fixed, by adding our delta term, which makes their solutions well-defined for all $\bf{x}$.  More important, is to note that replacing the indentity operator by the (general) $\be$ matrix provides a generalization of their solutions for all spin-values.
\subsection{Conclusions}
From our analysis, we have the following conclusions concerning the square-root operator:
\begin{enumerate}
\item In the simplest case, $\bf{A}=0$, the square-root operator has a representation as a nonlocal composite of three singularities. The particle component has two negative parts and one (hard core) positive part, while the antiparticle component has two positive parts and one (hard core) negative part. This effect is confined within a Compton wavelength such that, at the point of singularity, they cancel each other providing a finite result.  Furthermore, the operator looks like the identity outside a  Compton wavelength.  (Recall that the  experimental observation of three singularities in proton and neutron scattering experiments led to the quark model.)
\item A  constant magnetic field induces changes in both the mass and the shape of this extended object. It also  increases the number of singularities.   This suggests that the square-root operator represents a charge/mass density, for otherwise it could not be affected by a constant magnetic field.
\item The square-root operator is not physically the same as the Dirac operator despite the fact that they are related by a unitary transformation.  (We will discuss this point further in the next section.)
\end{enumerate}
\section{The Dirac Equation}
The first successful attempt to resolve the question of how best to handle the square-root equation was made by Dirac  in 1926 \cite{13}.  Dirac noticed that the Pauli matrices could be used to write 
${c^2}{{\mathbf{p}}^2} + {m^2}{c^4}{\text{ as }}{\left[ {c\alpha  \cdot {\mathbf{p}} + m{c^2}\beta } \right]^2}$.  The matrix $\al$ is defined by $\alpha  = \left( {{\alpha _1},{\alpha _2},{\alpha _3}} \right)$, where
\[
{\alpha _i} = \left( {\begin{array}{*{20}{c}}
  {\mathbf{0}}&{{\sigma _i}} \\ 
  {{\sigma _i}}&{\mathbf{0}} 
\end{array}} \right),\quad {\sigma _1} = \left( {\begin{array}{*{20}{c}}
  {\mathbf{0}}&1 \\ 
  1&{\mathbf{0}} 
\end{array}} \right),{\text{ }}{\sigma _2} = \left( {\begin{array}{*{20}{c}}
  {\mathbf{0}}&{ - i} \\ 
  i&{\mathbf{0}} 
\end{array}} \right),{\text{ }}{\sigma _3} = \left( {\begin{array}{*{20}{c}}
  {\mathbf{1}}&{\mathbf{0}} \\ 
  {\mathbf{0}}&{ - {\mathbf{1}}} 
\end{array}} \right).
\]
Thus, Dirac \cite{13} showed that an alternative representation of the square-root equation could be taken as:
\beqn
i\hbar \frac{{\partial \Psi }}{{\partial t}} = \left[ {c\alpha  \cdot {\mathbf{p}} + m{c^2}\beta } \right]\Psi .
\eeqn
In this case, $\Psi$ must be viewed as a vector-valued function or spinor.  To be more precise,
$\Psi  \in {L^2}\left( {{\mathbb{R}^3},{\mathbb{C}^4}} \right) = {L^2}\left( {{\mathbb{R}^3}} \right) \otimes {\mathbb{C}^4}$ is a four-component column vector $\Psi  = {\left( {{\psi _1},{\psi _2},{\varphi _1},{\varphi _2}} \right)^t}$.  In this representation, $\psi=({\psi _1},{\psi _2})^t$ represents the particle (positive energy) component, and $\varphi=({\varphi _1},{\varphi _2})$  represents the antiparticle (negative energy) component of the theory (for details, see Thaller \cite{14}).

A fair understanding of the Dirac equation can only be claimed in recent times, and, as pointed out by D. Finkelstein, ``Dirac introduced a Lorentz-invariant Clifford algebra into the complex algebra of observables of the electron".   (See, in particular, Biedenharn \cite{15} or deVries \cite{16} and Hestenes \cite{17}.)
Despite successes, both practical and theoretical, there still remain a number of conceptual, interpretational, and technical misunderstandings about this equation.  It is generally believed that it is not possible to separate the particle and antiparticle components directly without approximations (when interactions are present).  The various approximations found in the literature may have led to this belief.  In addition, the historically important algebraic approaches of Foldy-Wouthuysen \cite{18}, Pauli \cite{19}, and Feynman and Gell-Mann \cite{20} have no doubt further supported such ideas.

In this section we show that it is possible to directly separate the particle and antiparticle components of the Dirac equation without approximations, even when scalar and vector potentials of quite general character are present (see \cite{9}).  We show that the square root operator cannot be considered physically equivalent to the Dirac operator from another point of view. In addition, we offer another interpretation of the zitterbewegung and the fact that the expected value of a velocity measurement of a Dirac particle at any instant of time is $\pm c$.  
\subsection{Complete Separation}
It turns out that a direct analytic separation is actually quite simple and provides additional insight into the particle and antiparticle components.  In order to see this, let ${\mathbf{A}}({\mathbf{x}},t)$ and
$V({\mathbf{x}})$ be given vector and scalar potentials and, after adding $V({\mathbf{x}})$ and making the transformation ${\mathbf{p}} \to \pi  = {\mathbf{p}} - e{\kern 1pt} /{\kern 1pt} c{\mathbf{A}}$,  write (2.1) in two-component form as:
\beqn
\begin{gathered}
  i\hbar \frac{{\partial \psi }}{{\partial t}} = (V + m{c^2})\psi  + c(\sigma  \cdot \pi )\varphi  \hfill \\
  i\hbar \frac{{\partial \varphi }}{{\partial t}} = (V - m{c^2})\varphi  + c(\sigma  \cdot \pi )\psi . \hfill \\ 
\end{gathered} 
\eeqn
 
We write the second equation as:
\[
\left[ {\frac{\partial }{{\partial t}} + iB_1} \right]\varphi  = -iD\psi ,\quad B_1 = \tfrac{1}{\hbar }(V - m{c^2}){\text{  and  }}D = \tfrac{1}{\hbar }c(\sigma  \cdot \pi ).
\]
In this form, we see that from an analytical point of view equation (2.2) is a first order  inhomogeneous partial differential equation.   This equation can be solved via the Green's function method if we first solve 
\[
\left[ {\frac{\partial }{{\partial t}} + iB_1} \right]u(t) = \delta (t).
\]
It is easy to see that the solution to this equation is
\[
u(t) = \theta (t)\exp \{  - iB_1t\} ,{\text{   }}\theta (t) = \left\{ {\begin{array}{*{20}{c}}
  {1, \; t  > 0} \\ 
  {0, \; t < 0} 
\end{array}} \right. ,
\]
so that
\[
\varphi (t) = \int_{ - \infty }^t {c\exp \{  - iB_1(t - \tau )\} \left[ {{{(\sigma  \cdot \pi )} \mathord{\left/
 {\vphantom {{(\sigma  \cdot \pi )} {i\hbar }}} \right.
 \kern-\nulldelimiterspace} {i\hbar }}} \right]\psi (\tau )d\tau }. 
\]
It now follows via convolution that:
\beqn
\begin{gathered}
  i\hbar \frac{{\partial \psi }}{{\partial t}} = (V + m{c^2})\psi  \hfill \\
   + \left[ {{{{c^2}(\sigma  \cdot \pi )} \mathord{\left/
 {\vphantom {{{c^2}(\sigma  \cdot \pi )} {i\hbar }}} \right.
 \kern-\nulldelimiterspace} {i\hbar }}} \right]\int_{ - \infty }^t {\exp \{  - iB_1(t - \tau )\} (\sigma  \cdot \pi )\psi (\tau )d\tau }.  \hfill \\ 
\end{gathered} 
\eeqn
In a similar manner, we obtain the complete equation for $\varphi$:
\beqn
\begin{gathered}
  \hbar \frac{{\partial \varphi }}{{\partial t}} = (V - m{c^2})\varphi  \hfill \\
   + \left[ {{{{c^2}(\sigma  \cdot \pi )} \mathord{\left/
 {\vphantom {{{c^2}(\sigma  \cdot \pi )} {i\hbar }}} \right.
 \kern-\nulldelimiterspace} {i\hbar }}} \right]\int_{ - \infty }^t {\exp \{  - iB_2(t - \tau )\} (\sigma  \cdot \pi )\varphi (\tau )d\tau } , \hfill \\ 
\end{gathered} 
\eeqn
where
\[
v(t) = \theta (t)\exp \{  - iB_2t\} ,{\text{   }}\theta (t) = \left\{ {\begin{array}{*{20}{c}}
  {1, \; t  > 0} \\ 
  {0, \; t < 0} 
\end{array}} \right. ,
\]
$B_2=\tf{1}{\hbar}(V+mc^2)$.

Thus, we have decomposed ${L^2}\left( {{\mathbb{R}^3},{\mathbb{C}^4}} \right)$ as ${L^2}\left( {{\mathbb{R}^3},{\mathbb{C}^4}} \right) = {L^2}\left( {{\mathbb{R}^3},{\mathbb{C}^2}} \right) \oplus {L^2}\left( {{\mathbb{R}^3},{\mathbb{C}^2}} \right)$.  One copy of ${L^2}\left( {{\mathbb{R}^3},{\mathbb{C}^2}} \right)$ contains the particle (positive energy) wave component, while the other copy contains the antiparticle (negative energy) wave component.   Which of these copies corresponds to the components $\psi  = {\left( {{\psi _1},{\psi _2}} \right)^t}$  and which to the components   $\varphi  = {\left( {{\varphi _1},{\varphi _2}} \right)^t}$ depends, to some extent, on the properties of the scalar potential $V$.  It may have been noticed that equations (2.3) and (2.4) are non-hermitian.  It is shown in [9] that they are mapped into each other by the charge conjugation operator, so that the full matrix representation is hermitian.    An unsettled issue is the definition of the appropriate inner product for the two subspaces, which will account for the quantum constraint that the total probability integral is normalized.  We can satisfy this requirement if we set $\left( {\psi ,} \right.\left. \chi  \right) = {\psi _1}{{\bar \chi }_1} + {\psi _2}{{\bar \chi }_2}$,   $\left( {\psi ,} \right.{\left. \chi  \right)_1} = \left( {{A_1}\psi ,} \right.\left. {{A_1}\chi } \right)$ and $\left( {\varphi ,} \right.{\left. \eta  \right)_2} = \left( {{A_2}\varphi ,} \right.\left. {{A_2}\eta } \right)$, where ${A_1}\psi  = cu(t) * \left[ {{{(\sigma  \cdot \pi )} \mathord{\left/
 {\vphantom {{(\sigma  \cdot \pi )} {i\hbar }}} \right.
 \kern-\nulldelimiterspace} {i\hbar }}} \right]\psi (t)$ and ${A_2}\varphi  = cv(t) * \left[ {{{(\sigma  \cdot \pi )} \mathord{\left/
 {\vphantom {{(\sigma  \cdot \pi )} {i\hbar }}} \right.
 \kern-\nulldelimiterspace} {i\hbar }}} \right]\varphi (t)$.  We can now define the particle and antiparticle inner products by:
\beqn
\begin{gathered}
  \left\langle \psi  \right.,{\left. \chi  \right\rangle _p} = \int_{{{\mathbf{R}}^3}}^{} {\left[ {\left( {\psi ,} \right.\left. \chi  \right) + \left( {\psi ,} \right.{{\left. \chi  \right)}_1}} \right]d{\mathbf{x}}}  \hfill \\
  \left\langle \varphi  \right.,{\left. \eta  \right\rangle _{ap}} = \int_{{{\mathbf{R}}^3}}^{} {\left[ {\left( {\varphi ,} \right.\left. \eta  \right) + \left( {\varphi ,} \right.{{\left. \eta  \right)}_2}} \right]d{\mathbf{x}}},  \hfill \\ 
\end{gathered} 
\eeqn
so that the normalized probability densities satisfy:
\beqn
\begin{gathered}
  {\rho _\psi } = {\left| \psi  \right|^2} + {\left| {\int_{ - \infty }^t {c\exp \{  - iB(t - \tau )\} \left[ {{{(\sigma  \cdot \pi )} \mathord{\left/
 {\vphantom {{(\sigma  \cdot \pi )} {i\hbar }}} \right.
 \kern-\nulldelimiterspace} {i\hbar }}} \right]\psi (\tau )d\tau } } \right|^2} \hfill \\
  {\rho _\varphi } = {\left| \varphi  \right|^2} + {\left| {\int_{ - \infty }^t {c\exp \{  - iB'(t - \tau )\} \left[ {{{(\sigma  \cdot \pi )} \mathord{\left/
 {\vphantom {{(\sigma  \cdot \pi )} {i\hbar }}} \right.
 \kern-\nulldelimiterspace} {i\hbar }}} \right]\varphi (\tau )d\tau } } \right|^2}. \hfill \\ 
\end{gathered} 
\eeqn
\subsubsection{Interpretations}
Writing the Dirac equation and the direct separation in two-component matrix form, we have:
\[
i\hbar \frac{\partial }{{\partial t}}\left[ {\begin{array}{*{20}{c}}
  \psi  \\ 
  \varphi  
\end{array}} \right] = \left[ {\begin{array}{*{20}{c}}
  {(V + m{c^2})}&{c(\sigma  \cdot \pi )} \\ 
  {c(\sigma  \cdot \pi )}&{(V - m{c^2})} 
\end{array}} \right]\left[ {\begin{array}{*{20}{c}}
  \psi  \\ 
  \varphi  
\end{array}} \right]
\]
and
\[
i\hbar \frac{\partial }{{\partial t}}\left[ {\begin{array}{*{20}{c}}
  \psi  \\ 
  \varphi  
\end{array}} \right] = \left[ {\begin{array}{*{20}{c}}
  \begin{gathered}
  {\text{         }}(V + m{c^2}) \hfill \\
   + \left[ {{{{c^2}(\sigma  \cdot \pi )} \mathord{\left/
 {\vphantom {{{c^2}(\sigma  \cdot \pi )} {i\hbar }}} \right.
 \kern-\nulldelimiterspace} {i\hbar }}} \right]\left[ {u * (\sigma  \cdot \pi )} \right] \hfill \\ 
\end{gathered} &{\mathbf{0}} \\ 
  {\mathbf{0}}&\begin{gathered}
  {\text{           }}(V - m{c^2}) \hfill \\
   + \left[ {{{{c^2}(\sigma  \cdot \pi )} \mathord{\left/
 {\vphantom {{{c^2}(\sigma  \cdot \pi )} {i\hbar }}} \right.
 \kern-\nulldelimiterspace} {i\hbar }}} \right]\left[ {v * (\sigma  \cdot \pi )} \right] \hfill \\ 
\end{gathered}  
\end{array}} \right]\left[ {\begin{array}{*{20}{c}}
  \psi  \\ 
  \varphi  
\end{array}} \right]
\]
We call the latter equation the analytic diagonalization of the Dirac equation because the wave function has not changed.

The standard approach to the diagonalization of the Dirac equation (without an external potential $V$)  is via the Foldy-Wouthuysen representation \cite{18}.  Assuming that $\bf{A}$ does not depend on $t$, the following generalization can be found in deVries \cite{16}:
\[
i\hbar \frac{\partial }{{\partial t}}\left[ {\begin{array}{*{20}{c}}
  {{\Phi _1}} \\ 
  {{\Phi _2}} 
\end{array}} \right] = \left[ {\begin{array}{*{20}{c}}
  {\sqrt {{c^2}{\pi ^2} - ec\hbar (\Sigma  \cdot {\rm B}) + {m^2}{c^4}} }&0 \\ 
  0&{ - \sqrt {{c^2}{\pi ^2} - ec\hbar (\Sigma  \cdot {\rm B}) + {m^2}{c^4}} } 
\end{array}} \right]\left[ {\begin{array}{*{20}{c}}
  {{\Phi _1}} \\ 
  {{\Phi _2}} 
\end{array}} \right]
\]
where
\[
{\Sigma} = \left( {\begin{array}{*{20}{c}}
  {\bm{\sigma}}&{{\mathbf{0}}} \\ 
  {{\mathbf{0}}}&{\bm{\sigma}} 
\end{array}} \right).
\]
In this case,  ${\left[ {\begin{array}{*{20}{c}}
  {{\Phi _1}}&{{\Phi _2}} \end{array}} \right]^t} = {U_{FW}}{\left[ {\begin{array}{*{20}{c}}
  \psi &\varphi  \end{array}} \right]^t}$ and our square-root operator $S = {U_{FW}}{{\mathbf{H}}_{\text{D}}}U_{FW}^{ - 1}$.

From equation (2.2), we conclude that the coupling of the particle and antiparticle wave functions in the first-order form of the Dirac equation hides the second order nonlocal time nature of the equation.  We know that the square-root operator is nonlocal in space.  Thus, the implicit time nonlocality of the Dirac equation is mapped  into the explicit spatial nonlocality of the square-root equation by the Foldy-Wouthuysen transformation.  This is a mathematical relationship, which is not physically equivalent.

The time nonlocal behavior raises questions about the zitterbewegung.  The physically reasonable interpretation of the zitterbewegung and the fact that the expected value of a velocity measurement (of a Dirac particle) at any instant in time $\pm c$ are reflections of the fact that the Dirac equation makes a spatially extended particle appear as a point in the present by forcing it to oscillate between the past and future at speed  $\pm c$.  
\section{Classical Proper-time Theory}
In this section, we briefly review the classical theory.  The theory was first introduced in 2001 \cite{21} and further discussed in \cite{22}.  However, the theory has its roots in the foundations of quantum electrodynamics as developed by Feynman  and Dyson.
\subsection{Background}
Following the suggestions of Feynman and Dyson, our program began with the development of a mathematical theory for Feynman's time-ordered operator calculus, where time is accorded its natural role as the director of physical processes.   Briefly, our theory is constructive in that operators acting at different times actually commute (in the mathematical sense). This approach allows us to develop a general perturbation theory for all theories generated by unitary evolutions.  We are also able to reformulate our theory as a physically motivated sum over paths as suggested by Feynman. 
Our purpose was to prove the last two remaining conjectures of Dyson concerning the mathematical foundations for QED (see \cite{23}). (A. Salam confirmed Dyson's first conjecture \cite{24}, while S. Weinberg \cite{25} confirmed his second one.)  In particular, we showed that:
\begin{enumerate}
\item The renormalized perturbation series of quantum electrodynamics is at most asymptotic. (We also provided the remainder so that, in the mathematical sense, the expansion is exact.)
\item The ultraviolet divergence of quantum electrodynamics is caused by a violation of the time-energy uncertainty relations (at each point in time).
\end{enumerate}
{\it As} a special case, our approach also provided the first rigorous mathematical foundation for the Feynman path integral formulation of quantum mechanics (see \cite{8}). 

In the Feynman world-view the universe is a three-dimensional motion picture in which more and more of the future appears as time evolves. Time is a physically defined variable with properties distinct from those of the three spacial variables.  This view is inconsistent with the Minkowski world-view, in which time is an additional coordinate for space-time geometry.

With this inconsistency in mind, we began to investigate the possibility that an alternative formulation of both classical and quantum theory could exist, which encodes the Feynman world-view.   We discovered the canonical proper-time approach to classical electrodynamics, in which the proper-time of the observed system is used as opposed to the proper-time of the observer.  
\subsection{Maxwell's equations}
For the local-time version of Maxwell's equations, it is convenient to start with the standard definition of proper-time:
\beqn
d\tau ^2  = dt^2  - \frac{1}{{c^2 }}d{\mathbf{x}}^2  = dt^2 \left[ {1 - \frac{{{\mathbf{w}}^2 }}{{c^2 }}} \right],\quad {\mathbf{w}} = \frac{{d{\mathbf{x}}}}{{dt}}.
\eeqn
Motivated by geometry and the philosophy of the times, Minkowski suggested that we use the proper-time to define a metric for the space-time implementation of the special theory of relativity.      Physically, it is well-known that   $d{\tau}$ is not an exact one-form because a particle can traverse many different paths (in space) during any given $\tau$ interval.  This reflects the fact that the distance traveled in a given $\tau$ interval depends on the forces acting on the particle.  This also  implies that the clock of the source carries additional physical information about the acting forces.    In order to see this, rewrite equation (3.1) as:
\beqn
dt^2  = d\tau ^2  + \frac{1}{{c^2 }}d{\mathbf{x}}^2  = d\tau ^2 \left[ {1 + \frac{{{\mathbf{u}}^2 }}{{c^2 }}} \right],\quad {\mathbf{u}} = \frac{{d{\mathbf{x}}}}
{{d\tau }}.
\eeqn
For any other observer, we have:
\beqn
dt'^2  = d\tau ^2  + \frac{1}{{c^2 }}d{\mathbf{x}}'^2  = d\tau ^2 \left[ {1 + \frac{{{\mathbf{u}}'^2 }}{{c^2 }}} \right],\quad {\mathbf{u}}' = \frac{{d{\mathbf{x}}'}}
{{d\tau }}.
\eeqn
It follows that observers can use one unique clock to discuss all events associated with the source (simultaneity). We also note that, the phase space variables remain unchanged because the momentum ${\mathbf{p}} = m{\mathbf{w}} = m_0 {\mathbf{u}}$, where $m = \gamma m_0 $.  

From equations (3.2) and (3.3) we see explicitly that, the new metric for each observer is exact, while the representation space is now  Euclidean.  In order to clearly see that we have a change in the clock convention, assume that we are observing a particle moving with constant velocity relative to our unprimed (inertial) frame.  In this case, we can integrate equation (3.2) obtaining:
\[
t = \left( {\sqrt {1 + \tfrac{{{{\mathbf{u}}^2}}}{{{c^2}}}} } \right)\tau. 
\]
The inverse relationship is
\[
\tau  = {\left( {\sqrt {1 + \tfrac{{{{\mathbf{u}}^2}}}{{{c^2}}}} } \right)^{ - 1}}t = \left( {\sqrt {1 - \tfrac{{{{\mathbf{w}}^2}}}{{{c^2}}}} } \right)t,
\]
where $\bf{w} =\tf{d\bf{x}}{dt}$.  Now, $t$ and $\tau$ differ by a scale factor, so that either may be used (a convention).  The advantage of the $\tau$ representation is that the same $\tau$ is also available to our prime observer in her frame:
\[
\tau  = {\left( {\sqrt {1 + \tfrac{{{{{\mathbf{u'}}}^2}}}{{{c^2}}}} } \right)^{ - 1}}t' = \left( {\sqrt {1 - \tfrac{{{{{\mathbf{w'}}}^2}}}{{{c^2}}}} } \right)t'.
\]
The important point of our theory is that, this convention is available in the non constant velocity case.   (In order to show that general case does not complicate matters, in Section 3.3 we construct the transformation group for all cases.) 

In our new formalism, the natural definition of velocity is no longer ${\bf{w}}=d{\bf{x}}/dt$ but ${\bf{u}}=d{\bf{x}}/d{\tau}$. This suggests that there may be a certain duality in the relationship between $t,\;\tau$ and ${\bf{w}},\;{\bf{u}}$.  To see that this is indeed the case, recall that ${\mathbf{u}} = {{\mathbf{w}} \mathord{\left/
 {\vphantom {{\mathbf{w}} {\sqrt {1 - \left( {{{{\mathbf{w}}^2 } \mathord{\left/
 {\vphantom {{{\mathbf{w}}^2 } {c^2 }}} \right. \kern-\nulldelimiterspace} {c^2 }}} \right)} }}} \right. \kern-\nulldelimiterspace} {\sqrt {1 - \left( {{{{\mathbf{w}}^2 } \mathord{\left/
 {\vphantom {{{\mathbf{w}}^2 } {c^2 }}} \right. \kern-\nulldelimiterspace} {c^2 }}} \right)} }}$.  Solving for ${\bf{w}}$, we get that ${\mathbf{w}} = {{\mathbf{u}} \mathord{\left/ {\vphantom {{\mathbf{u}} {\sqrt {1 + \left( {{{{\mathbf{u}}^2 } \mathord{\left/ {\vphantom {{{\mathbf{u}}^2 } {c^2 }}} \right. \kern-\nulldelimiterspace} {c^2 }}} \right)} }}} \right. \kern-\nulldelimiterspace} {\sqrt {1 + \left( {{{{\mathbf{u}}^2 } \mathord{\left/ {\vphantom {{{\mathbf{u}}^2 } {c^2 }}} \right. \kern-\nulldelimiterspace} {c^2 }}} \right)} }}$.    If we set $b = \sqrt {c^2  + {\mathbf{u}}^2 }$, this relationship can be written as  
\beqn      
\frac{{\mathbf{w}}}{c} = \frac{{\mathbf{u}}}{b}.
\eeqn
For reasons to be clear momentarily, we call $b$ the collaborative speed of light.  Indeed, we see that 
\beqn
\frac{1}{c}\frac{\partial }{{\partial t}} = \frac{1}{c}\frac{{\partial \tau }}
{{\partial t}}\frac{\partial }{{\partial \tau }} = \frac{1}{c}\frac{1}
{{\sqrt {1 + \left( {{{{\mathbf{u}}^2 } \mathord{\left/{\vphantom {{{\mathbf{u}}^2 } {c^2 }}} \right.\kern-\nulldelimiterspace} {c^2 }}} \right)} }}\frac{\partial }{{\partial \tau }} = \frac{1}{b}\frac{\partial }{{\partial \tau }}.
\eeqn
For our prime observer, it is easy to see that the corresponding result will be:
\beqn
\frac{{{\bf{w}}'}}
{c} = \frac{{{\bf{u}}'}}
{{b'}},\quad \quad \frac{1}
{c}\frac{\partial }
{{\partial t'}} = \frac{1}
{{b'}}\frac{\partial }
{{\partial \tau }}.
\eeqn
From equations (3.5) (and (3.6)) we see  that the non-invariance of $t, (t')$ and the invariance of $c$ on the left is replaced by the non-invariance of $b, (b')$ and the invariance of $\tau$ on the right.  These equations  represent mathematically equivalent relations. Thus, wherever they are used consistently as replacements for each other, they can't change the mathematical relationships. In order to see their impact on Maxwell's equations, in c.g.s. units, we have:
\beqn
\begin{gathered}
\nabla  \cdot {\mathbf{B}} = 0,\quad \quad \quad \nabla  \cdot {\mathbf{E}} = 4\pi \rho , \hfill \\
\nabla  \times {\mathbf{E}} =  - \frac{1}{c}\frac{{\partial {\mathbf{B}}}}{{\partial t}},\quad \nabla  \times {\mathbf{B}} = \frac{1}{c}\left[ {\frac{{\partial {\mathbf{E}}}}
{{\partial t}} + 4\pi \rho {\mathbf{w}}} \right]. \hfill \\ 
\end{gathered} 
\eeqn
Using equations (3.1) and (3.2) in (3.3), we have {{\it{the mathematically identical representation for Maxwell's equations}}:
\beqa
\begin{gathered}
\nabla  \cdot {\mathbf{B}} = 0,\quad \quad \quad \nabla  \cdot {\mathbf{E}} = 4\pi \rho , \hfill \\
\nabla  \times {\mathbf{E}} =  - \frac{1}{b}\frac{{\partial {\mathbf{B}}}}{{\partial \tau }},\quad \nabla  \times {\mathbf{B}} = \frac{1}{b}\left[ {\frac{{\partial {\mathbf{E}}}}
{{\partial \tau }} + 4\pi \rho {\mathbf{u}}} \right]. \hfill \\ 
\end{gathered} 
\eeqa
Thus, Maxwell's equations are equally valid when the local time of the particle is used to describe the fields.  This leads to the following conclusions:
\begin{enumerate}
\item There are two distinct clocks to use in the representation of Maxwell's equations.  (The choice of clocks is a convention.) 
\item Since the two representations are mathematically equivalent, mathematical equivalence is not the same as physical equivalence.      
\item When the proper-time is used, the constant speed of light $c$ is replaced by the effective speed of light $b$, which depends on the motion of the system (i.e., $b = \sqrt {c^2  + {\mathbf{u}}^2 }$).  Thus, we have a natural varying speed of light theory (VSL), as opposed to a postulated one ( see Magueijo \cite{26} or Moffat \cite{27}). 
\end{enumerate}
Let us now derive the corresponding wave equations in the local-time variable.  Taking the curl of the last two equations Maxwell equations (above), and using standard vector identities, we get: 
\beqn
\begin{gathered}
 \frac{1}{{b^2 }}\frac{{\partial^2 {\mathbf{B}}}}{{\partial \tau ^2 }} - \frac{{{\mathbf{u}} \cdot {\mathbf{a}}}}{{b^4 }}\left[ {\frac{{\partial {\mathbf{B}}}}
{{\partial \tau }}} \right] - \nabla ^2 \cdot {\mathbf{B}} = \frac{1}{b}\left[ 4\pi \nabla  \times (\rho {\mathbf{u}}) \right], \hfill \\
 \frac{1}{{b^2 }}\frac{{\partial ^2 {\mathbf{E}}}}{{\partial \tau ^2 }} - \frac{{{\mathbf{u}} \cdot {\mathbf{a}}}}{{b^4 }}\left[ {\frac{{\partial {\mathbf{E}}}}
{{\partial \tau }}} \right] - \nabla ^2  \cdot {\mathbf{E}} =  - \nabla (4\pi \rho ) - \frac{1}{b}\frac{\partial }{{\partial \tau }}\left[ {\frac{{4\pi (\rho {\mathbf{u}})}}{b}} \right]. \hfill \\ 
\end{gathered} 
\eeqn				
where ${\bf{a}} = d{\bf{u}}/d\tau$ is the effective acceleration.  The new (gauge independent) term appears instantaneously along the direction of motion, but opposing  any applied force and is zero otherwise.  This is the near field (i.e., the field at the site of the charged particle).  This is exactly what one expects of the back reaction caused by the inertial resistance of a particle to accelerated motion and, according to Wheeler and Feynman \cite{28}, is precisely what is meant by radiation reaction. Thus, the collaborative use of the observer's coordinate system and the local clock of the observed system provides intrinsic  information about the local field dynamics not available in the conventional formulation of Maxwell's theory.  It is shown in \cite{22}, that the theory does not require point particles, self-energy divergence,  mass renormalization or the Lorentz Dirac equation.     

It is also shown in  \cite{21} that, for a closed system of interacting charged particles, the proper-time of the center of mass corresponds to the historical clock of Horwitz, Piron, and Fanchi (see \cite{29} and  \cite{30}).  In this case, $b=c$ and the corresponding Maxwell equations represent the far field (only retarded potentials).  It was further shown that, from this vantage point, the particle interactions appear as the delayed action-at-a-distance type.  This verifies the Wheeler-Feynman conjecture that field theory and delayed action-at-a-distance are complimentary manifestations of the same physics (see \cite{28}).  The requirement of total conservation of momentum, angular momentum and energy allowed us to prove the complete absorption of radiation by all particles in the system.    (Recall that, this was an assumption in the Wheeler and Feynman approach {\it and the center-fold of their theory}.)

If we make a scale transformation (at fixed position) with ${\bf{E}} \to (b/c)^{1/2}{\bf{E}}$    and ${\bf{B}} \to (b/c)^{1/2}{\bf{B}}$, the equations in (3.4) transform to 
\beqn
\begin{gathered}
 \frac{1}{{b^2 }}\frac{{\partial ^2 {\mathbf{B}}}}{{\partial \tau ^2 }} - {\text{ }}\nabla ^2 {\kern 1pt}  \cdot {\mathbf{B}} + \left[ \frac{{\ddot b}}
{{2b^3 }}-{\frac{{3\dot b^2 }}{{4b^4 }}  } \right]{\mathbf{B}} = \frac{{c^{1/2} }}{{b^{3/2} }}\left[ {4\pi \nabla  \times (\rho {\mathbf{u}})} \right], \hfill \\
 \frac{1}{{b^2 }}\frac{{\partial ^2 {\mathbf{E}}}}{{\partial \tau ^2 }} - {\text{ }}\nabla ^2 {\kern 1pt}  \cdot {\mathbf{E}} +  \left[ \frac{{\ddot b}}
{{2b^3 }}-{\frac{{3\dot b^2 }}{{4b^4 }}  } \right]{\mathbf{E}} =  - \frac{{c^{1/2} }}{{b^{1/2} }}\nabla (4\pi \rho ) - \frac{{c^{1/2} }}{{b^{3/2}}}\frac{\partial}{{\partial \tau }}\left[ {\frac{{4\pi (\rho {\mathbf{u}})}}{b}} \right]. \hfill \\ 
\end{gathered} 
\eeqn
This is the Klein-Gordon equation with an effective mass $\mu$ given by
\beqn
\quad \quad \mu  = \left\{ {\frac{{\hbar ^2 }}{{c^2 }}\left[ {\frac{{\ddot b}}
{{2b^3 }} - \frac{{3\dot b^2 }}{{4b^4 }}} \right]} \right\}^{1/2}  = \left\{ {\frac{{\hbar ^2 }}{{c^2 }}\left[ {\frac{{{\mathbf{u}} \cdot {\mathbf{\ddot u}} + {\mathbf{\dot u}}^2 }}  {{2b^4 }} - \frac{{5\left( {{\mathbf{u}} \cdot {\mathbf{\dot u}}} \right)^2 }}
{{4b^6 }}} \right]} \right\}^{1/2}. 
\eeqn
\begin{rem} We note that, when $b$ is constant, $ \bf{a}=0, \; \mu=0$ and $t=\tf{b}{c} \tau$ (also $t'=\tf{b'}{c} \tau$), so that the local time theory is both mathematically and physically equivalent to the standard theory.  However, when $b$ is not constant $\mu \ne 0$ and the two approaches are not physically equivalent. 
\end{rem}
For additional insight, let $({\bf{x}}(\tau),\tau)$ represent the field position and $(\bar{\bf{x}}(\tau'),\tau')$ the retarded position of a source charge $e$, with ${\bf r}={ \bf{x}}-\bar{\bf{x}}$.  If we set $r=\left|{ \bf{x}}-\bar{\bf{x}}\right|$, $s=r-(\tfrac{({\bf r}\cdot{\bf u})}{b})$, and ${\bf r}_{\bf u}={\bf r}-{\tfrac{r}{b}}{\bf u}$, then we were able to compute the $\bf E$ and $\bf B$ fields directly in [21]  to obtain:
\beqa
\quad \quad {\mathbf{E}}({\mathbf{x}},\tau ) = \frac{{e\left[ {{\mathbf{r}}_{\mathbf{u}} (1 - {{{\mathbf{u}}^2 } \mathord{\left/
 {\vphantom {{{\mathbf{u}}^2 } {b^2 }}} \right.
 \kern-\nulldelimiterspace} {b^2 }})} \right]}}
{{s^3 }} + \frac{{e\left[ {{\mathbf{r}} \times ({\mathbf{r}}_{\mathbf{u}}  \times {\mathbf{a}})} \right]}}
{{b^2 s^3 }} + \frac{{e({\mathbf{u}} \cdot {\mathbf{a}})\left[ {{\mathbf{r}} \times ({\mathbf{u}} \times {\mathbf{r}})} \right]}}
{{b^4 s^3 }}
\eeqa
and
\[
{\mathbf{B}}({\mathbf{x}},\tau ) = \frac{{e\left[ {({\mathbf{r}} \times {\mathbf{r}}_{\mathbf{u}} )(1 - {{{\mathbf{u}}^2 } \mathord{\left/
 {\vphantom {{{\mathbf{u}}^2 } {b^2 }}} \right.
 \kern-\nulldelimiterspace} {b^2 }})} \right]}}
{{rs^3 }} + \frac{{e{\mathbf{r}} \times \left[ {{\mathbf{r}} \times ({\mathbf{r}}_{\mathbf{u}}  \times {\mathbf{a}})} \right]}}
{{rb^2 s^3 }} + \frac{{er({\mathbf{u}} \cdot {\mathbf{a}})({\mathbf{r}} \times {\mathbf{u}})}}
{{b^4 s^3 }}.
\]
(It is easy to see that $\bf B$ is orthogonal to $\bf E$.)  The first two terms in the above equations are standard, in the $({\bf{x}}(t),{\bf{w}}(t))$ variables.  The third part of both equations is new and arises because of the dissipative term in our wave equation.  (Once again, this term is zero when $b$ is constant.)  It is easy to see that ${\bf r}\times({\bf u}\times{\bf r})=r^2{\bf u}-({\bf u}\cdot{\bf r}){\bf r}$, so we get a component along the direction of motion. (Thus,  the $\bf E$ field has a  longitudinal part.) This confirms our claim that the new dissipative term is equivalent to an effective mass that arises due to the collaborative acceleration of the particle.   This means that the cause for radiation reaction comes directly from the use of the local clock to formulate Maxwell's equations.  It follows that, in this  approach there is no need to assume advanced potentials, self-interaction, mass renormalization and the Lorentz-Dirac equation in order to account for it (radiation reaction), as is required when the observer clock is used. Furthermore, no assumptions about the structure of the source are needed.  

\subsection{\bf Proper-time Lorentz Group} 
We now identify the new transformation group that preserves the first postulate of the special theory. The standard (Lorentz) time transformations between two inertial observers can be written as
\beqn
t' = \gamma ({\mathbf{v}})\left[ {t - {{{\mathbf{x}} \cdot {\mathbf{v}}} \mathord{\left/ {\vphantom {{{\mathbf{x}} \cdot {\mathbf{v}}} {c^2 }}} \right.
 \kern-\nulldelimiterspace} {c^2 }}} \right], \quad \quad \quad \quad {\text{      }}t = \gamma ({\mathbf{v}})\left[ {t' + {{{\mathbf{x'}} \cdot {\mathbf{v}}} \mathord{\left/
 {\vphantom {{{\mathbf{x'}} \cdot {\mathbf{v}}} {c^2 }}} \right. \kern-\nulldelimiterspace} {c^2 }}} \right].
\eeqn
We want to replace $t,\;t'$ by $\tau$.  To do this, use the relationship between $dt$ and $d\tau$ to get:
\beqn
t = \tfrac{1}{c}\int_0^\tau  {b(s)} ds = \tfrac{1}{c}\bar b\tau ,\quad  t' = \tfrac{1}{c}\int_0^\tau  {b'(s)} ds = \tfrac{1}{c}\bar b'\tau, 
\eeqn
where we have used the mean value theorem of calculus to obtain the end result, so that both $\bar b$ and $\bar b'$ represent an earlier $\tau$-value of $b$ and $b'$ respectively.  Note that, as $b$ and $b'$ depend on $\tau$, the transformations (3.8) represent explicit nonlinear relationships between $t, t'$ and $\tau$ (during interaction).  This is to be expected in the general case when the system is acted on by external forces.   However, if $b$ is constant (so is $b'$), then $t, \ t'$ and $\tau$ differ by a scale transformation, which means they are physically equivalent, in addition to their natural mathematical equivalence.

If we set
\[
{\mathbf{d}}^ *   = {{\mathbf{d}} \mathord{\left/ {\vphantom {{\mathbf{d}} {\gamma ({\mathbf{v}})}}} \right.\kern-\nulldelimiterspace} {\gamma ({\mathbf{v}})}} - (1 - \gamma ({\mathbf{v}}))\left[ {({{{\mathbf{v}} \cdot {\mathbf{d}})} \mathord{\left/
{\vphantom {{{\mathbf{v}} \cdot {\mathbf{d}})} {(\gamma ({\mathbf{v}}){\mathbf{v}}^2 }}} \right. \kern-\nulldelimiterspace} {(\gamma ({\mathbf{v}}){\mathbf{v}}^2 }})} \right]{\mathbf{v}},
\] 
we can write the transformations that fix $\tau$ as:
\beqn
\begin{gathered}
\quad {\mathbf{x'}} = \gamma ({\mathbf{v}})\left[ {{\mathbf{x}}^ *   - ({{\mathbf{v}} \mathord{\left/{\vphantom {{\mathbf{v}} c}} \right.\kern-\nulldelimiterspace} c})\bar b\tau } \right],\quad \quad \quad \quad \,{\mathbf{x}} = \gamma ({\mathbf{v}})\left[ {{\mathbf{x'}}^*   + ({{\mathbf{v}} \mathord{\left/{\vphantom {{\mathbf{v}} c}} \right.\kern-\nulldelimiterspace} c})\bar b'\tau } \right], \hfill \\
\quad \quad \quad\quad {\mathbf{u'}} = \gamma ({\mathbf{v}})\left[ {{\mathbf{u}}^ *   - ({{\mathbf{v}} \mathord{\left/{\vphantom {{\mathbf{v}} c}} \right.\kern-\nulldelimiterspace} c})b} \right],\quad\quad\quad    \; \,{\text{   }}{\mathbf{u}} = \gamma ({\mathbf{v}})\left[ {{\mathbf{u'}}^*   + ({{\mathbf{v}} \mathord{\left/ {\vphantom {{\mathbf{v}} c}} \right. \kern-\nulldelimiterspace} c})b'} \right], \hfill \\
\quad {\mathbf{a'}} = \gamma ({\mathbf{v}})\left\{ {{\mathbf{a}}^*  - {\mathbf{v}}\left[ {({{{\mathbf{u}} \cdot {\mathbf{a}})} \mathord{\left/{\vphantom {{{\mathbf{u}} \cdot {\mathbf{a}})} {(bc}}} \right.\kern-\nulldelimiterspace} {(bc}})} \right]} \right\},\quad {\text{   }}{\mathbf{a}} = \gamma ({\mathbf{v}})\left\{ {{\mathbf{a'}}^*   + {\mathbf{v}}\left[ {({{{\mathbf{u'}} \cdot {\mathbf{a'}})} \mathord{\left/{\vphantom {{{\mathbf{u'}} \cdot {\mathbf{a'}})} {(b'c}}} \right. \kern-\nulldelimiterspace} {(b'c}})} \right]} \right\}. \hfill \\ 
\end{gathered} 
\eeqn
If we put equation (3.8) in (3.7), differentiate with respect to $\tau$  and cancel the extra factor of $c$, we get the transformations between $b$ and $b'$:
\beqn
\quad\quad\quad \quad b'(\tau ) = \gamma ({\mathbf{v}})\left[ {b(\tau ) - {{{\mathbf{u}} \cdot {\mathbf{v}}} \mathord{\left/{\vphantom {{{\mathbf{u}} \cdot {\mathbf{v}}} c}} \right. \kern-\nulldelimiterspace} c}} \right],\quad \quad b(\tau ) = \gamma ({\mathbf{v}})\left[ {b'(\tau ) + {{{\mathbf{u'}} \cdot {\mathbf{v}}} \mathord{\left/
 {\vphantom {{{\mathbf{u'}} \cdot {\mathbf{v}}} c}} \right. \kern-\nulldelimiterspace} c}} \right].
\eeqn
From these results, it follows that, at the local level, during interaction, equations (3.9) and (3.10) provide a nonlinear and nonlocal representation of the Lorentz group. We call it the proper-time Lorentz group.

\subsection{Proper-Time Particle Theory}
We now investigate the corresponding particle theory.   The key concept to our approach may be seen by examining the time evolution of a dynamical parameter $W({\bf{x}},{\bf{p}})$, via the standard formulation of classical mechanics, described in terms of the Poisson brackets:
\beqn
\frac{{dW}}{{dt}} = \left\{ {H,W} \right\}.
\eeqn
We can also represent the dynamics via the proper time by using the representation $d\tau  = ({1 \mathord{\left/{\vphantom {1 \gamma }} \right. \kern-\nulldelimiterspace} \gamma })dt = ({{mc^2 } \mathord{\left/ {\vphantom {{mc^2 } H}} \right. \kern-\nulldelimiterspace} H})dt$,
so that:
\[
\frac{{dW}}{{d\tau }} = \frac{{dt}}{{d\tau }}\frac{{dW}}{{dt}} = \frac{H}
{{mc^2 }}\left\{ {H,W} \right\}.
\]
Assuming a well-defined (invariant) rest energy ($mc^2$) for the particle, we determine the canonical proper-time Hamiltonian $K$ such that:
\[
\left\{ {K,W} \right\} = \frac{H}{{mc^2 }}\left\{ {H,W} \right\},\quad \left. K \right|_{{\mathbf{p}} = 0}  = \left. H \right|_{{\mathbf{p}} = 0}  = mc^2. 
\]
Using 
\[
\begin{gathered}
  \left\{ {K,W} \right\} = \left[ {\frac{H}{{mc^2 }}\frac{{\partial H}}{{\partial {\mathbf{p}}}}} \right]\frac{{\partial W}}{{\partial {\mathbf{x}}}} - \left[ {\frac{H}
{{mc^2 }}\frac{{\partial H}}{{\partial {\mathbf{x}}}}} \right]\frac{{\partial W}}
{{\partial {\mathbf{p}}}} \hfill \\
  {\text{          }} = \frac{\partial }{{\partial {\mathbf{p}}}}\left[ {\frac{{H^2 }}
{{2mc^2 }} + a} \right]\frac{{\partial W}}{{\partial {\mathbf{x}}}} - \frac{\partial }
{{\partial {\mathbf{x}}}}\left[ {\frac{{H^2 }}{{2mc^2 }} + a'} \right]\frac{{\partial W}}
{{\partial {\mathbf{p}}}}, \hfill \\ 
\end{gathered} 
\]
we get that $a = a' = \tfrac{1}{2}mc^2$, so that (assuming no explicit time dependence)
\[
K = \frac{{H^2 }}{{2mc^2 }} + \frac{{mc^2 }}{2},\quad {\text{and }}\quad \frac{{dW}}
{{d\tau }} = \left\{ {K,W} \right\}.
\] 
Since $\tau$ is invariant during interaction (minimal coupling), we make the natural assumption that (the form of) $K$ also remains invariant.  Thus, if $\sqrt {c^2 {\mathbf{p}}^2  + m^2 c^4 }  \to \sqrt {c^2 {\bs{\pi}}^2  + m^2 c^4 }  + V$, where  $\bf A$ a vector potential, $V$ is a potential energy term and  $\pi  = {\mathbf{p}} - \tfrac{e}
{c}{\mathbf{A}}$.  In this case, $K$ becomes:
\[
K=\frac{{ {\bs{\pi }}^2}}
{{2m}} + mc^2  + \frac{{V^2 }}
{{2mc^2 }} + \frac{{V\sqrt {c^2  {\bs{\pi }}^2+ m^2 c^4 } }}
{{mc^2 }}.
\]
If we set $H_0=\sqrt {c^2 {\bs{\pi }}^2  + m^2 c^4 }$, use standard vector identities with $H_0=mcb -V$, $\nabla \times \bs\pi=-\tfrac{e}{c}\bf{B}$, and compute Hamilton's equations, we get:
\[
{\mathbf{u}} = \frac{{d{\mathbf{x}}}}{{d\tau }} = \left[ {1 + \frac{V}{{{H_0}}}} \right]\frac{{{\bs{\pi} }}}{m} = \left[ {\frac{{mbc}}{{mbc - V}}} \right]\frac{{\bs{\pi }}}{m} \Rightarrow {\bs{\pi }} = m{\mathbf{u}} - \frac{V}{{bc}}{\mathbf{u}}
 \]
and
\beqn
\begin{gathered}
  \frac{{d{\bf{p}}}}
{{d\tau }} =  - \frac{{\left[ {\left( {\bs \pi  \cdot \nabla } \right) \bs \pi  + \tfrac{e}
{c}\bs \pi  \times {\bf{B}}} \right]}}
{m}\left[ {1 + \frac{V}
{{H_0 }}} \right] - \nabla V\frac{{H_0 }}
{{mc^2 }}\left[ {1 + \frac{V}
{{H_0}}} \right] \hfill \\
   \quad \quad  = \tfrac{e}
{c}  \left( {{\bf{u}} \cdot \nabla } \right) {\bf{A}}  + \tfrac{e}
{c}{\bf{u}} \times {\bf{B}} - \nabla V \f{b}{c}\left[ {1 + \frac{V}
{{mcb }}} \right]. \hfill \\ 
\end{gathered} 
\eeqn
Further reduction, using the definition of $\bf{E}$, with $V = e\Phi $, we have:
\beqn 
 \begin{gathered}
  \frac{c}
{b}\left[ {\frac{{d{\mathbf{p}}}}
{{d\tau }} - \frac{e}
{c}\frac{{d{\mathbf{A}}}}
{{d\tau }}} \right] =  - \frac{e}
{b}\frac{{\partial {\mathbf{A}}}}
{{\partial \tau }} + \tfrac{e}
{b}{\mathbf{u}} \times {\mathbf{B}} - e\nabla \Phi \left[ {1 + \frac{V}
{{mcb }}} \right] \hfill \\
  \quad \quad \quad \quad \quad \quad  = e{\mathbf{E}} + \tfrac{e}
{b}{\mathbf{u}} \times {\mathbf{B}} -e \nabla \Phi \frac{V}
{{mcb }}. \hfill \\ 
\end{gathered}
\eeqn
It is clear that the additional term in equation (3.7) acts to oppose the force imposed by the charged particle part of the $\bf{E}$ field (i.e., $- \nabla V$).   In order to see the physical meaning of the term, assume an interaction between a proton and an electron, where $\bf{A}=0$ and $V$ is the Coulomb interaction, so that (3.7)  becomes:
\beqn 
  \frac{c}
{b}\frac{{d{\mathbf{p}}}}
{{d\tau }} =  - \nabla V - \nabla V\frac{V}
{{mcb }}. 
\eeqn
Using $H_0 \approx mc^2$, we see that $\lim_{r \to r_0}\bf{u} =0$ and $\lim_{r \to r_0}\bf{a} =0$, so that:
\beqn
0 =  - \nabla V - \nabla V\frac{V}
{{mc^2 }}
\eeqn
and the classical electron radius, $r_0$, is a critical point (i.e., $-\nabla V   -\nabla V (V/mc^2)=0$).  Thus, for $0< r< r_0$, the force becomes repulsive.  We interpret this as a fixed region of repulsion, so that the singularity $r=0$ is impossible to reach at the classical level.  The neglected terms are attractive but of lower order.  This makes the critical point less than $r_0$.  Thus, in general, the electron experiences a strongly repulsive force when it gets too close to the proton.  This means that the classical principle of impenetrability, namely that no two particles can occupy the same space at the same time occurs naturally.   It is this additional term that leads us to suspect that the electron may not act like a point particle in the s-states of hydrogen, where it has a finite probability of being at the center of the proton. 

The above observation also implies that, two electrons will experience an attraction if they can come close enough together as for example, at very low energies (temperatures).

The Lagrangian representation reveals the close relationship to the non-relativistic case.  If we solve for $\bf{p}$, we get 
\[
  {\mathbf{p}} = m{\mathbf{u}} - \frac{{V{\mathbf{u}}}}{{cb}} + \frac{e}{c}{\mathbf{A}}.
\]
Using this in $K$ along with  $b^2 =  {{u^2} + {c^2}}$, we have
\[
\begin{gathered}
  K = \frac{{{{\left( {m{\mathbf{u}} - \frac{{V{\mathbf{u}}}}{{bc}}} \right)}^2}}}{{2m}} + m{c^2} + \frac{{{V^2}}}{{2m{c^2}}} + \frac{{V\left( {mcb - V} \right)}}{{m{c^2}}} \hfill \\
   = \tfrac{1}{2}m{u^2} - \frac{{V{u^2}}}{{bc}} + \frac{{{V^2}{u^2}}}{{2m{b^2}{c^2}}} + m{c^2} - \frac{{{V^2}}}{{2m{c^2}}} + \frac{{Vb}}{c}. \hfill \\ 
\end{gathered} 
\]
From $\mcL d\tau  = {\bf{p}} \cdot d{\bf{x}} - Kd\tau $, we can write $\mcL$ as
\[
\begin{gathered}
  \mcL = \left[ {m{\mathbf{u}} - \frac{{V{\mathbf{u}}}}{{bc}}} \right] \cdot {\mathbf{u}} + \tfrac{e}{c}{\mathbf{A}} \cdot {\mathbf{u}} \hfill \\
   - \left\{ {\tfrac{1}{2}m{u^2} - \frac{{V{u^2}}}{{bc}} + \frac{{{V^2}{u^2}}}{{2m{b^2}{c^2}}} + m{c^2} + \frac{{Vb}}{c} - \frac{{{V^2}}}{{2m{c^2}}}} \right\} \hfill \\
= \tfrac{1}{2}m{u^2} + \tfrac{e}{c}{\mathbf{A}} \cdot {\mathbf{u}} - m{c^2} - \frac{{Vb}}{c} + \frac{{{V^2}}}{{2m{c^2}}}\left[ {1 - \frac{{{u^2}}}{{{b^2}}}} \right]. \hfill \\ 
\end{gathered} 
\]
From this representation, it is clear that the neglect of second order terms gives us the non-relativistic theory. 

\section{{Relativistic Quantum Theory}}
The Klein-Gordon and Dirac equations were first discovered in early  attempts to make quantum mechanics compatible with the Minkowski formulation of special theory of relativity.   Both were partially successful but could no longer be interpreted as particle equations and a complete theory required quantum fields and the associated problems.  For a recent discussion of other problems, one can consult \cite{9} (see also \cite{6}).  
	
In this section we introduce the canonical extension of the Dirac and square-root equations.  	
Let ${\mathbf{A}}({\mathbf{x}},t)$ and $V({\mathbf{x}})$ be given vector and scalar potentials and, after adding $V({\mathbf{x}})$ and making the transformation ${\mathbf{p}} \to \pi  = {\mathbf{p}} - \tfrac{e}{c}{\mathbf{A}}$.  

To quantize our theory, we follow the standard procedure leading to the equation:
\[
i\hbar \frac{{\partial \Phi }}{{\partial \tau }} = K\Phi  = \left[ {\frac{{{H^2}}}{{2m{c^2}}} + \frac{{m{c^2}}}{2}} \right]\Phi .
\]
However, in addition to the Dirac Hamiltonian, there are two other possible Hamiltonians, depending on the way the potential appears with the square-root operator:
\[
{\beta} \sqrt {{c^2}\bs{\pi}^2  - ec\hbar \Sigma  \cdot {\mathbf{B}} + {m^2}{c^4}}  + V
\]
and
\[
{\beta} \sqrt {{c^2}\bs{\pi}^2  - ec\hbar \Sigma  \cdot {\mathbf{B}} + {{\left( {m{c^2} + {\beta} V} \right)}^2}}. 
\] 
We have identified three possible canonical proper-time particle equations for spin-$\tfrac{1}{2}$ particles. (We also note that, these equations can be modified to apply to particles of any spin, by a minor change in the $\be$ matrix.)  
\begin{enumerate}
\item The canonical proper-time version of the Dirac equation:  
\beqn
\begin{gathered}
  i\hbar \frac{{\partial \Psi }}{{\partial \tau }} = \left\{ {\frac{{{\bs{\pi}^2}}}{{2m}} + \beta {V} + m{c^2}} \right. - \frac{{e\hbar \Sigma  \cdot {\mathbf{B}}}}{{2mc}} \hfill \\
  \quad \quad \;\;\left. { + \frac{{{V}\alpha  \cdot \bs{\pi} }}{{mc}} - \frac{{i\hbar \alpha  \cdot \nabla {V}}}{{2mc}} + \frac{{{V^2}}}{{2{mc^2}}}} \right\}\Psi . \hfill \\ 
\end{gathered} 
\eeqn
\item The canonical proper-time version of the square-root equation, using the first possibility:
\beqn
\begin{gathered}
i\hbar \frac{{\partial \Psi }}{{\partial \tau}} = \left\{ \frac{{\bs{\pi}^2 }}{{2m}} - \frac{{e\hbar \Sigma  \cdot {\bf{B}}}}{{2mc }} + mc^2  + \frac{{V^2 }}{{2mc^2 }}\right\}\Psi  \hfill \\
+ \frac{{V\beta \sqrt {c^2 \bs{\pi}^2  - ec\hbar \Sigma  \cdot {\bf{B}} + m^2 c^4 } }}{{2mc^2 }}\Psi
+ \frac{{\beta \sqrt {c^2 \bs{\pi}^2  - ec\hbar \Sigma  \cdot {\bf{B}} + m^2 c^4 } }}{{2mc^2 }}V \Psi. \hfill \\
\end{gathered} 
\eeqn
\item The canonical proper-time version of the square-root equation, using the second possibility:
\beqn
  i\hbar \frac{{\partial \Psi }}{{\partial \tau }} = \frac{{{\bs{\pi}^2}}}{{2m}} + {\beta}V + m{c^2} - \frac{{e\hbar \Sigma  \cdot {\mathbf{B}}}}{{2mc}} + \frac{{{V^2}}}{{2m{c^2}}}.  
\eeqn
\end{enumerate} 
If $V=0$, all equations reduce to:
\[
i\hbar \frac{{\partial \Psi }}{{\partial \tau }} = \left\{ {\frac{{{\bs{\pi}^2}}}{{2m}} + m{c^2}} \right. - \left. {\frac{{e\hbar \Sigma  \cdot {\mathbf{B}}}}{{2mc}}} \right\}\Psi .
\]
The close relationship to the Schr\"{o}dinger operator, makes it easy to see that, in all cases, $K$ is positive definite.  In mathematical terms, the lower order terms are relatively bounded with respect to ${\bs{\pi}^2}/{2m}$.  It follows that, unlike the Dirac  and Klein-Gordon approach, we can interpret (4.1)-(4.3) as true particle equations. In the above equations, we have assumed that $V$ is time independent.  (However,  since ${\mathbf{A}}({\mathbf{x}},t)$ can have general time-dependence,  $\sqrt {c^2 \bs{\pi}^2  - ec\hbar \Sigma  \cdot {\bf{B}} + m^2 c^4 }$ need not be related to the Dirac operator by a Foldy-Wouthuysen type transformation.)

We plan to investigate the last two equations at a later time.  In the next section, we focus on the canonical proper-time Dirac extension.
\subsection{The Dirac Theory}
Since the Dirac equation forms the basis for QED,  an important test of our proper-time extension is how well it compares to the Dirac equation in its description of the hydrogen spectrum.  In this section, we compare the  Dirac equation with the canonical proper-time extension for the Hydrogen atom problem. 
 
If we let $\bf{A}=\bf{0}$, $V_0=-\tf{e^2}{r}$ and consider the standard Dirac Hydrogen atom eigenvalue problem,
\[
{\lambda _n}{\Psi _n} = {H_D}{\Psi _n},
\]
where $\la_n$ is the $n$-th eigenvalue and  ${\Psi _n}$ is the  corresponding eigenfunction.  For this case, if $j$ is the total angular momentum and $\al$ is the fine structure constant, we have
\beqa
{\lambda _n} = m{c^2}\left[ {1 + \frac{{{\alpha ^2}}}{{{{\left[ {n - \left| {j + \tfrac{1}{2}} \right| + \sqrt {{{\left( {j + \tfrac{1}{2}} \right)}^2} - {\alpha ^2}} } \right]}^2}}}} \right]^{-1/2}.
\eeqa
For the proper-time extension, with the same eigenfunction, we have
\[
{E_n}{\Psi _n} = \left[ {\frac{{H_D^2}}{{2m{c^2}}} + \frac{{m{c^2}}}{2}} \right]{\Psi _n} = \left[ {\frac{{\lambda _n^2}}{{2m{c^2}}} + \frac{{m{c^2}}}{2}} \right]{\Psi _n}.
\]
In order to compare the two expressions to order $O(\al^8)$, we begin with the following approximations:
\[
\begin{gathered}
(1) \quad  {\left( {1 + x} \right)^{ - 1}} \simeq 1 + x + {x^2} + O({x^3}) \hfill \\
(2) \quad  {\left( {1 + x} \right)^{ - 1/2}} \simeq 1 - \tfrac{1}{2}x + \tfrac{3}{8}{x^2} + O({x^3}) \hfill \\
(3) \quad  {\left( {1 - x} \right)^{1/2}} \simeq 1 - \tfrac{1}{2}x - \tfrac{1}{8}{x^2} + O({x^3}). \hfill \\ 
\end{gathered} 
\]
Let $\kappa ={j + \tfrac{1}{2}}$, then  our proper time extension can be written as:
\beqn
\begin{gathered}
  {E_n} = \left[ {\frac{{\lambda _n^2}}{{2m{c^2}}} + \frac{{m{c^2}}}{2}} \right] = \frac{{m{c^2}}}{2}\left\{ {1 + {{\left[ {1 + \tfrac{{{\alpha ^2}}}{{{{\left( {n - \left| \kappa  \right| + \sqrt {{\kappa ^2} - {\alpha ^2}} } \right)}^2}}}} \right]}^{ - 1}}} \right\} \hfill \\
   \simeq \frac{{m{c^2}}}{2}\left\{ {1 + \left[ {1 - \tfrac{{{\alpha ^2}}}{{{{\left( {n - \left| \kappa  \right| + \sqrt {{\kappa ^2} - {\alpha ^2}} } \right)}^2}}} + \tfrac{{{\alpha ^4}}}{{{{\left( {n - \left| \kappa  \right| + \sqrt {{\kappa ^2} - {\alpha ^2}} } \right)}^4}}}} \right]} \right\} \hfill \\
   = \frac{{m{c^2}}}{2}\left[ {2 - \tfrac{{{\alpha ^2}}}{{{{\left( {n - \left| \kappa  \right| + \sqrt {{\kappa ^2} - {\alpha ^2}} } \right)}^2}}} + \tfrac{{{\alpha ^4}}}{{{{\left( {n - \left| \kappa  \right| + \sqrt {{\kappa ^2} - {\alpha ^2}} } \right)}^4}}}} \right]. \hfill \\ 
\end{gathered} 
\eeqn
Using (2), we have:
\[
\begin{gathered}
  {\lambda _n} \simeq m{c^2}{\left[ {1 +\tfrac{{{\alpha ^2}}}{{{{\left( {n - \left| \kappa  \right| + \sqrt {{\kappa ^2} - {\alpha ^2}} } \right)}^2}}}} \right]^{ - 1/2}} \hfill \\
   \simeq m{c^2}\left[ {1 - \tfrac{{{\alpha ^2}}}{{2{{\left( {n - \left| \kappa  \right| + \sqrt {{\kappa ^2} - {\alpha ^2}} } \right)}^2}}} + \tfrac{{3{\alpha ^4}}}{{8{{\left( {n - \left| \kappa  \right| + \sqrt {{\kappa ^2} - {\alpha ^2}} } \right)}^4}}}} \right]. \hfill \\ 
\end{gathered} 
\]
Using (3), we can approximate  $\sqrt{{\kappa ^2} - {\alpha ^2}}$ to get
\[
\begin{gathered}
  \sqrt {{\kappa ^2} - {\alpha ^2}}  \simeq \left| \kappa  \right|\left( {1 - \tfrac{{{\alpha ^2}}}{{2{\kappa ^2}}} - \tfrac{{{\alpha ^4}}}{{8{\kappa ^4}}}} \right) = \left| \kappa  \right| - \tfrac{{{\alpha ^2}}}{{2\left| \kappa  \right|}} \Rightarrow  \hfill \\
  n - \left| \kappa  \right| + \sqrt {{\kappa ^2} - {\alpha ^2}}  \simeq n - \tfrac{{{\alpha ^2}}}{{2\left| \kappa  \right|}} \Rightarrow  \hfill \\
  {\alpha ^2}{\left[ {n - \left| \kappa  \right| + \sqrt {{\kappa ^2} - {\alpha ^2}} } \right]^{ - 2}} \simeq \frac{{{\alpha ^2}}}{{{n^2}}}{\left[ {{{\left( {1 - \tfrac{{{\alpha ^2}}}{{2n\left| \kappa  \right|}}} \right)}^2}} \right]^{ - 1}} \simeq \frac{{{\alpha ^2}}}{{{n^2}}}\left[ {1 + \tfrac{{{\alpha ^2}}}{{n\left| \kappa  \right|}} - \tfrac{{{\alpha ^4}}}{{4{n^2}{{\left| \kappa  \right|}^3}}}} \right] \hfill \\
   = \frac{{{\alpha ^2}}}{{{n^2}}} + \frac{{{\alpha ^4}}}{{{n^3}\left| \kappa  \right|}} - \frac{{{\alpha ^6}}}{{4{n^4}{{\left| \kappa  \right|}^3}}} \hfill \\ 
\end{gathered} 
\]
and 
\[
{\alpha ^4}{\left[ {n - \left| \kappa  \right| + \sqrt {{\kappa ^2} - {\alpha ^2}} } \right]^{ - 4}} \simeq \frac{{{\alpha ^4}}}{{{n^4}}}{\left[ {{{\left( {1 - \tfrac{{{\alpha ^2}}}{{2n\left| \kappa  \right|}}} \right)}^4}} \right]^{ - 1}} \simeq \frac{{{\alpha ^4}}}{{{n^4}}}\left[ {1 + \tfrac{{2{\alpha ^2}}}{{n\left| \kappa  \right|}}} \right].
\]
With the last result, we now have:
\beqn
\begin{gathered}
  {\lambda _n} \simeq m{c^2}\left\{ {1 - \frac{{{\alpha ^2}}}{{2{n^2}}}\left[ {1 + \frac{{{\alpha ^2}}}{{n\left| \kappa  \right|}} - \frac{{{\alpha ^4}}}{{4n{{\left| \kappa  \right|}^3}}}} \right] + \frac{{3{\alpha ^4}}}{{8{n^4}}}\left[ {1 + \frac{{{\alpha ^2}}}{{n\left| \kappa  \right|}} } \right]} \right\} \hfill \\
  = m{c^2}\left\{ {\left[ {1 - \frac{{{\alpha ^2}}}{{2{n^2}}} - \frac{{{\alpha ^4}}}{{2{n^4}}}\left( {\frac{n}{{\left| \kappa  \right|}} - \frac{3}{4}} \right)} \right] + \frac{{{\alpha ^6}}}{{8{n^5}{\left| \kappa  \right|}}}\left( {\frac{n^2}{{\left| \kappa  \right|^2}} + {3}} \right)} \right\}. \hfill \\ 
\end{gathered} 
\eeqn
For $E_n$, we have
\beqn
\begin{gathered}
  {E_n} \simeq \frac{{m{c^2}}}{2}\left\{ {2 - \left[ {\frac{{{\alpha ^2}}}{{{n^2}}} + \frac{{{\alpha ^4}}}{{{n^3}\left| \kappa  \right|}} - \frac{{{\alpha ^6}}}{{4{n^4}{{\left| \kappa  \right|}^2}}}} \right] + \frac{{{\alpha ^4}}}{{{n^4}}}\left[ {1 + \frac{{2{\alpha ^2}}}{{n\left| \kappa  \right|}}} \right]} \right\} \hfill \\
   = m{c^2}\left\{ {\left[ {1 - \frac{{{\alpha ^2}}}{{2{n^2}}} - \frac{{{\alpha ^4}}}{{2{n^4}}}\left( {\frac{n}{{\left| \kappa  \right|}} - 1} \right)} \right] + \frac{{{\alpha ^6}}}{{4{n^5}\left| \kappa  \right|}}\left( {\frac{n}{{\left| \kappa  \right|}} + 8} \right)} \right\}. \hfill \\ 
\end{gathered}
\eeqn
It is now easy to see that, to order $\al^4, \; \la_n-E_n=\tf{-\al^4}{8n^4}$, so that the $E_n$ values are systematically lower than the $\la_n$ values.

Table 1 below provides a relative comparison between the Dirac and proper-time extension compared with the experimental data for s-states, compiled by National Institute for Standards and Technology (NIST) of the US government. 

{ \bf Table 1: Comparison with NIST data for  s-states}

		\begin{tabular}{|l||l|c|l|p{3.0cm}|l|} 
		\hline
		 State & Dirac & Proper-time & Nist & $\De$-DNIST & $\De$-PTNIST\\
		\hline\hline
		2s  & 10.20439429 & 10.20422448 & 10.19881008  & .00558421  & .00541440\\
		\hline
		3s  & 12.09411035 & 12.09393146 & 12.08749443  & .00661592 & .00643603\\
		\hline
		4s  & 12.75550914 & 12.75532871 & 12.74853244  & .00697670 & .00679627\\
		\hline
		5s  & 13.06164150 & 13.06146066 & 13.05449789   & .00714361  & .00696277\\
	\hline
		\end{tabular} 
\newline
\newline
As can be seen from the last two columns, the proper-time extension consistently provides results that a closer to the experimental data for all cases.  (We could not compare the two 1s-states with experiment, because of the NIST normalization for this state.)
  
In Table 2, we see the same comparative results for the p, d and f-states. 

{ \bf Table 2: Comparison with NIST data for p, d and f-states}
\smallskip

		\begin{tabular}{|l||l|c|l|p{3.0cm}|l|} 
		\hline
		 State & Dirac & Proper-time & Nist & $\De$-DNIST & $\De$-PTNIST\\
		\hline\hline
		2p (j=1/2) & 10.20439429 & 10.20422448 & 10.19880553 & 0.005588760 & 0.005418952\\
		\hline
		2p (j=3/2) & 10.20443957 & 10.20426976 & 10.19885089 & 0.005588681  & 0.005418870\\
		\hline
		3p (j=1/2) & 12.09411035 & 12.09393146 & 12.08749292 & 0.006617431  & 0.006438537\\
		\hline
		3p (j=3/2) & 12.09412377 & 12.09394488 & 12.08750636 & 0.006617407  & 0.006438512\\
		\hline
		3d (j=3/2) & 12.09412377 & 12.09394488 & 12.08750634 & 0.006617430  & 0.006438535\\
		\hline
		3d (j=5/2) & 12.09412824 & 12.09394935 & 12.08751082 & 0.006617422  & 0.006438528\\
		\hline
		4p (j=1/2) & 12.75550914 & 12.75532871 & 12.74853167 & 0.006977467  & 0.006797044\\
		\hline
		4f (j=7/2) & 12.75551763 & 12.75533720 & 12.74854038 &  0.006976250 & 0.006796820 \\

	\hline
		\end{tabular}

Thus, in all cases, the canonical proper-time extension of the Dirac equation provides a closer approximation to the known experimental data for the Hydrogen spectra compared to the Dirac equation.  In all cases, the changes are in the forth decimal place.  This is insufficient to account for either the Lamb shift or the anomalous magnetic moment.  

\subsection{Future Direction}
In what follows, let $V=V_0=\tf{-e^2}{r}$.  Based on our analysis of the square-root operator in the first section and the Dirac operator in the second section, we are in the process of investigating the possibility that in s-states, the potential energy takes on the form:
\beqn
\begin{gathered}
  V =  - \frac{{{e^2}\sqrt {{M^2}{c^4} - ec\hbar \Sigma  \cdot {\mathbf{B}} + {c^2}{\pi ^2}} }}{{M{c^2}r}} =  - \frac{{{e^2}}}{r}\frac{{\left( {1 - \tfrac{{e\hbar \Sigma  \cdot {\mathbf{B}}}}{{{M^2}{c^3}}} + \tfrac{{{\pi ^2}}}{{{M^2}{c^2}}}} \right)}}{{\sqrt {1 - \tfrac{{e\hbar \Sigma  \cdot {\mathbf{B}}}}{{{M^2}{c^3}}} + \tfrac{{{\pi ^2}}}{{{M^2}{c^2}}}} }} \hfill \\
   \simeq {V_0} + \frac{{{r_0}}}{r}\frac{{e\hbar \Sigma  \cdot {\mathbf{B}}}}{{2Mc}} - \frac{{{r_0}}}{r}\frac{{{\pi ^2}}}{{2M}}, \hfill \\ 
\end{gathered} 
\eeqn
where $r_0=e^2/Mc^2$.  There are three possible choices for $M$: 
\begin{enumerate}
\item The electron  cannot be treated as a point particle in s-states of hydrogen, so that $M=m$, the mass of the electron and $r_0$ is the classical electron radius.
\item Neither the electron nor the proton can be treated as point particles in s-states of hydrogen, so that $M= \mu$, the reduced mass and $r_0$ is the classical mixed reduced radius.
\item The electron can be treated as a point particle in s-states of hydrogen, but the proton cannot so that $M=m_p$, the mass of the proton and $r_0$ is the classical reduced proton radius.
\end{enumerate}
It is clear that, at the zero-th order, we recover the Coulomb potential and the second term in (4.7)  is a first order approximation.   Assuming the first case, our eigenvalue problem becomes:
\beqn
\begin{gathered}
E \Psi  = \left\{ {\frac{{{\bs{\pi}^2}}}{{2m}} + \beta {V} + m{c^2}} \right. - \frac{{e\hbar \Sigma  \cdot {\mathbf{B}}}}{{2mc}} \hfill \\
  \quad \quad \;\;\left. { + \frac{{{V}\alpha  \cdot \bs{\pi} }}{{mc}} - \frac{{i\hbar \alpha  \cdot \nabla {V}}}{{2mc}} + \frac{{{V^2}}}{{2{mc^2}}}} \right\}\Psi . \hfill \\ 
\end{gathered} 
\eeqn
As a first try, we set $V=V_0$ for the terms containing $\al$, use our first order approximation in the second term and $V$ itself in the last term, so that
\beqa
 \be V   \simeq \be{V_0} + \be \frac{{{r_0}}}{r}\frac{{e\hbar \Sigma  \cdot {\mathbf{B}}}}{{2mc}} - \be \frac{{{r_0}}}{r}\frac{{{\pi ^2}}}{{2m}}.
\eeqa
For the last term, we use the approximation:
\[
\begin{gathered}
  \frac{{{V^2}}}{{2m{c^2}}} = \frac{1}{2}\left\{ {\frac{{{V_0}}}{{m{c^2}}}\sqrt {1 - \frac{{e\hbar \Sigma  \cdot {\mathbf{B}}}}{{{m^2}{c^3}}} + \frac{{{\pi ^2}}}{{{m^2}{c^2}}}} \frac{{{V_0}}}{{m{c^2}}}\sqrt {1 - \frac{{e\hbar \Sigma  \cdot {\mathbf{B}}}}{{{m^2}{c^3}}} + \frac{{{\pi ^2}}}{{{m^2}{c^2}}}} } \right\} \hfill \\
   \simeq \frac{1}{2}\left\{ {{{\left[ {\frac{{{V_0}}}{{m{c^2}}}} \right]}^2}\left[ {1 - \frac{{e\hbar \Sigma  \cdot {\mathbf{B}}}}{{{m^2}{c^3}}} + \frac{{{\pi ^2}}}{{{m^2}{c^2}}}} \right]} \right\} + {\frac{{{V_0}}}{{m{c^2}}}}\frac{{{{\mathbf{p}}^2}\left[ {{V_0}} \right]}}{{2{m^2}{c^2}}}\sqrt {1 - \frac{{e\hbar \Sigma  \cdot {\mathbf{B}}}}{{{m^2}{c^3}}} + \frac{{{\pi ^2}}}{{{m^2}{c^2}}}}  \hfill \\
   \simeq \frac{1}{2}\left\{ {{{\left[ {\frac{{{V_0}}}{{m{c^2}}}} \right]}^2}\left[ {1 - \frac{{e\hbar \Sigma  \cdot {\mathbf{B}}}}{{{m^2}{c^3}}} + \frac{{{\pi ^2}}}{{{m^2}{c^2}}}} \right]} \right\} +{\frac{{{V_0}}}{{m{c^2}}}} \frac{{{{\mathbf{p}}^2}\left[ {{V_0}} \right]}}{{2{m^2}{c^2}}} \hfill \\
   = \frac{1}{2}\left\{ {{{\left[ {\frac{{{V_0}}}{{m{c^2}}}} \right]}^2}\left[ {1 - \frac{{e\hbar \Sigma  \cdot {\mathbf{B}}}}{{{m^2}{c^3}}} + \frac{{{\pi ^2}}}{{{m^2}{c^2}}}} \right]} \right\} - {\frac{{{V_0}}}{{m{c^2}}}}\frac{{2\pi {\hbar ^2}}}{{{m^2}{c^2}}}\delta \left( r \right). \hfill \\ 
\end{gathered} 
\]
Using these terms, we have:
\[
\begin{gathered}
  E\Psi  = \left\{ {1 - \beta \frac{{{r_0}}}{r} + \frac{{r_0^2}}{{{r^2}}}} \right\}\frac{{{\pi ^2}}}{{2m}}\Psi  + \left[ {1 + \beta \frac{{{r_0}}}{r} - \frac{{r_0^2}}{{{r^2}}}} \right]\frac{{e\hbar \Sigma  \cdot {\mathbf{B}}}}{{2mc}}\Psi  + m{c^2}\Psi  \hfill \\
   + \beta {V_0} + \frac{{{V_0}\alpha  \cdot \pi }}{{mc}}\Psi  - \frac{{i\hbar \alpha  \cdot \nabla {V_0}}}{{2mc}}\Psi  + \frac{{V_0^2}}{{2m{c^2}}}\Psi  + {\frac{{{r_0}}}{{r}}}\frac{{2\pi {\hbar ^2}}}{{m{c^2}}}\delta \left( r \right)\Psi  \hfill \\
   = \left\{ {\frac{{{\pi ^2}}}{{2m}} + \be V_0- \frac{{e\hbar \Sigma  \cdot {\mathbf{B}}}}{{2mc}} + m{c^2} + \frac{{{V_0}\alpha  \cdot \pi }}{{mc}} - \frac{{i\hbar \alpha  \cdot \nabla {V_0}}}{{2mc}} + \frac{{V_0^2}}{{2m{c^2}}}} \right\}\Psi  \hfill \\
   + \left\{ {\left[ {\frac{{r_0^2}}{{{r^2}}} - \beta \frac{{{r_0}}}{r}} \right]\frac{{{\pi ^2}}}{{2m}} - \left[ {\frac{{r_0^2}}{{{r^2}}} - \beta \frac{{{r_0}}}{r}} \right]\frac{{e\hbar \Sigma  \cdot {\mathbf{B}}}}{{2mc}} + {\frac{{{r_0}}}{{r}}}\frac{{2\pi {\hbar ^2}}}{{m^2{c^2}}}\delta \left( r \right)} \right\}\Psi  \hfill \\
   = {E_0}\Psi  + K'\Psi , \hfill \\ 
\end{gathered} 
\]
where
\[
K' =  {\left[ {\frac{{r_0^2}}{{{r^2}}} - \beta \frac{{{r_0}}}{r}} \right]\frac{{{\pi ^2}}}{{2m}} - \left[ {\frac{{r_0^2}}{{{r^2}}} - \beta \frac{{{r_0}}}{r}} \right]\frac{{e\hbar \Sigma  \cdot {\mathbf{B}}}}{{2mc}} + {\frac{{{r_0}}}{{r}}}\frac{{2\pi {\hbar ^2}}}{{{m^2}{c^2}}}\delta \left( r \right)}.
\]
In conclusion, if this approach is as successful as we believe, we will still need to justify our approximation methods.

\subsection*{Acknowledgments}
The work reported here represents the outcome of a program begun by the author, J. Lindesay and W. W. Zachary at Howard University over thirty years ago.  The work in section two is joint with M. Alfred (Howard) and that in section three is joint with Gonzalo Ares de Parga (Mexico).

\bibliographystyle{amsalpha}

\end{document}